\documentclass{emulateapj} 
\usepackage{apjfonts}
\usepackage{epsfig}
\usepackage{xspace}
\usepackage{natbib}
\usepackage{hhline}
\usepackage{amsmath}
\usepackage{multirow}
\usepackage{dcolumn}
\usepackage{verbatim}
\usepackage[figuresright]{rotating}
\usepackage{calc}
\usepackage{dcolumn}

\newcolumntype{.}{D{.}{.}{-1}}

\newcommand{\chandra}{{\it Chandra}\xspace}
\newcommand{\suzaku}{{\it Suzaku}\xspace}
\newcommand{\HST}{{\it HST}\xspace}
\newcommand{\swift}{{\it Swift}\xspace}

\newcommand{\integral}{{\it INTEGRAL}\xspace}
\newcommand{\champlane}{ChaMPlane\xspace}

\newcommand{\rms}{{\it rms}\xspace}

\newcommand{\sS}[1]{\mbox{$\rm{}^{#1}$}}
\newcommand{\Ss}[1]{\mbox{$\rm{}_{#1}$}}

\newcommand{\Ms}{\mbox{$M_{\odot}$}\xspace}

\newcommand{\nHt}{\mbox{$N_{\mbox{\scriptsize H22}}$}\xspace}

\newcommand{\Deg}{\mbox{$^\circ$}\xspace}
\newcommand{\x}{\mbox{$\times$}}

\newcommand{\fap}{$P\mbox{$\rm{}_{FAP}$}$\xspace}
\newcommand{\detp}{$P\mbox{$\rm{}_{det}$}$\xspace}

\newcommand{\Ps}{\mbox{$P_{\mbox{\scriptsize s}}$}\xspace}
\newcommand{\Po}{\mbox{$P_{\mbox{\scriptsize o}}$}\xspace}

\begin{document}

\title{Discovery of a Significant Magnetic CV Population
in the Limiting Window}



\author{
JaeSub Hong\altaffilmark{1*}, 
Maureen van den Berg\altaffilmark{2},
Jonathan E. Grindlay\altaffilmark{1}, \\
Mathieu Servillat\altaffilmark{1}, 
and 
Ping Zhao\altaffilmark{1}
}
\altaffiltext{*}{Send requests to J. Hong at jaesub@head.cfa.harvard.edu}
\altaffiltext{1}{Harvard-Smithsonian Center for Astrophysics, 
60 Garden St., Cambridge, MA 02138 }
\altaffiltext{2}{Utrecht University, 3508 TC Utrecht, The Netherlands}

\begin{abstract} 

We have discovered 10 periodic X-ray sources from the 1 Ms \chandra
ACIS observation of the Limiting Window (LW),
a low extinction region ($A_V\sim 3.9$) at 1.4\Deg south of the Galactic
center. The LW provides a rare opportunity of studying the Galactic Bulge
sources without obscuration from molecular clouds. 
Using the early 100 ks \chandra exposure
and the \HST observation that covers five of these sources,
we have reported three of them as candidate CVs or accreting
binaries, based on their blue optical colors, excess H$\alpha$ fluxes,
and high X-ray-to-optical
flux ratios \citep{Berg09}. Our observations of the LW using the IMACS
camera on the Magellan 6.5 m telescope show one or
two candidate counterparts within the error circle of six sources,
and all show relatively high X-ray-to-optical flux ratios.
The observed periods ($\sim$ 1.3 to 3.4 hours) and the X-ray luminosities
($10^{31.8-32.9}$ erg
s\sS{-1} at 8 kpc) of the 10 periodic sources,
combined with the lack of bright optical counterparts and thus high
X-ray-to-optical flux ratios, suggest that they are likely accreting
binaries, in particular, magnetic cataclysmic variables (MCVs).  All of
the 10 sources exhibit a relatively hard X-ray spectrum ($\Gamma<2$ for
a power law model) and X-ray spectra of at least five show an extinction 
larger than the field average expected from the interstellar medium in
the region.
The latter implies some intrinsic absorption in the system, which is
also a typical sign of MCVs.  
The discovery of these periodic X-ray sources in the LW further supports
the current view that MCVs constitute the majority of low
luminosity hard X-ray sources ($\sim10^{30-33}$ erg s\sS{-1}) found in the 
Bulge.
The period distribution of these sources resembles those of polars,
whereas the relatively hard
spectra suggest that they could be intermediate polars (IPs).  
These puzzling properties, which are also shared by some of the periodic X-ray
sources found in the Sgr A* field \citep{Muno03b}, can be explained by
unusual polars with buried magnetic fields or a rare sub-class of MCVs,
nearly synchronous MCVs. These unusual MCVs may provide important clues
in the evolutionary path of MCVs from IPs to polars.  While the
\chandra X-ray band appears relatively more sensitive in discovering
these unusual MCVs,  the completeness simulation indicates $\gtrsim$40\% of
the hard X-ray sources in the LW are periodic.  Therefore, this discovery
provides a first direct evidence of a large MCV population in the Bulge.

\end{abstract}
\keywords{Galaxy: bulge --- X-ray: binaries --- cataclysmic variables}

\section{Introduction}

The high sensitivity and superb spatial resolution of 
\chandra enabled population study of low-luminosity
X-ray sources ($L_X\sim10^{30-34}$ erg s$^{-1}$) on Galactic
scales beyond the local solar neighborhood.  Several ongoing
campaigns, including our own
{\em Chandra} Multi-wavelength Plane (ChaMPlane) survey
\citep{Grindlay05} aim to improve the census of the Galactic
low-luminosity X-ray sources. 
The Galactic Bulge, in particular, has been of great interest.
More than 3000 discrete X-ray sources
have been discovered in the 17\arcmin \x 17\arcmin\ region around Sgr
A* \citep[hereafter M03a, M09]{Muno03a,Muno09}.  In ChaMPlane, we
study the Bulge through our
dedicated surveys of low-extinction bulge regions (``Windows survey'')
and a latitudinal strip around the Galactic center (``Bulge Latitude
Survey''), where thousands of X-ray sources have
been also discovered \citep[hereafter H09b]{Grindlay11,Hong09b}.
Multi-wavelength follow-up of these X-ray sources has been ongoing for
source classification through optical/infrared imaging and spectroscopy
\citep[e.g.][hereafter B06, B09]{Koenig08,Berg06,Berg09}.

H09b (see also M03a) have found that the X-ray sources in the
Galactic center region (GCR) show largely homogeneous X-ray properties
(e.g.~intrinsically hard X-ray spectra, photon power law index $\Gamma<$1
for a power law model). The lack of bright IR (K$<$15) counterparts for
the GCR X-ray sources indicates that HXMBs, once considered as a major
constitute for Bulge X-ray sources, cannot account
for more than 10\% of the population \citep[hereafter L05]{Laycock05}.  Currently the
leading candidates that fit the observed properties are magnetic
cataclysmic variables (MCVs) (L05; M09; H09b).  

\begin{table*}
\small
\caption{\chandra ACIS-I observations of the LW}
\begin{tabular*}{0.99\textwidth}{c@{\extracolsep{\fill}}ccccrrc}
\hline\hline
Obs.~ID	& Start Time		&R.A.	& Decl.		& Roll	&GTI\sS{a}	& Exposure	& Epoch\sS{b}	\\
 	& (UT: y/m/d h:m)	&(\Deg)	& (\Deg)	& (\Deg)&(ks)		& (ks)		& (duration)\\
\hline
6362	&2005/08/19 16:15	&267.86875 &-29.58800	& 273	&37.4		& 38 		\\
5934	&2005/08/22 08:16	&267.86875 &-29.58800	& 273	&36.1		& 41 		\\
\multicolumn{5}{r}{\it subtotal}				&\it 73.5	& \it 79 	&\it 1 (3.13 d) \\
\hline
6365	&2005/10/25 14:55	&267.86875 &-29.58800	& 265	&20.4		& 21 		\\
\multicolumn{5}{r}{\it subtotal}				&\it 20.4	& \it 21 	&\it 2 (0.24 d)\\
\hline
9505	&2008/05/07 15:29	&267.86375 &-29.58475	& 82	&10.7		& 11 		\\
9855	&2008/05/08 05:00	&267.86375 &-29.58475	& 82	&55.3		& 57 		\\
\multicolumn{5}{r}{\it subtotal}				&\it 66.0	& \it 68 	&\it 3 (1.22 d)\\
\hline
9502	&2008/07/17 15:45	&267.86375 &-29.58475	& 281	&161.7		& 167 		\\
9500	&2008/07/20 08:11	&267.86375 &-29.58475	& 280	&158.6		& 165 		\\
9501	&2008/07/23 08:13	&267.86375 &-29.58475	& 279	&130.0		& 135 		\\
9854	&2008/07/27 05:53	&267.86375 &-29.58475	& 278	&22.4		& 25 		\\
9503	&2008/07/28 17:37	&267.86375 &-29.58475	& 275	&99.5		& 103 		\\
9892	&2008/07/31 08:07	&267.86375 &-29.58475	& 275	&64.8		& 65 		\\
9893	&2008/08/01 02:44	&267.86375 &-29.58475	& 275	&41.4		& 45 		\\
9504	&2008/08/02 21:23	&267.86375 &-29.58475	& 275	&122.8		& 127 		\\
\multicolumn{5}{r}{\it subtotal}				&\it 801.2	& \it 832 	&\it 4 (17.7 d)\\
\hline
\end{tabular*}
\sS{a}The selected Good Time Interval (GTI) that is based on the lack 
of high background fluctuations ($<3\sigma$). See \citet{Hong05}.
\sS{b}The observations are grouped into four separate epochs (durations
given in days), so that the total span of each epoch does not exceed a month. 
\label{t:obs}
\end{table*}

Recently we started a search for periodic ChaMPlane sources
\citep[e.g.][hereafter H09a]{Hong09a}, in part, to circumvent difficulties in source
identification, which arise from large distances, high extinction (except
for the Windows fields), and source confusion due to high stellar density
in the Bulge.  In this work, we report the discovery of 10 periodic
X-ray sources from the 1 Ms \chandra
exposure of the Limiting Window (LW), a low extinction region at 1.4\Deg
south of the Galactic center
(see also H09b).

The low extinction Window fields, including the LW, provide a rare opportunity
of studying the GCR source population without obscuration from
molecular clouds
(H09a,H09b,B06,B09).  \citet[hereafter R09]{Revnivtsev09} 
showed that the Galactic Ridge X-ray Emission (GRXE), the nature
of which has been puzzling for decades, is mainly made up of discrete
faint sources of known nature, primarily active binaries (ABs) or
cataclysmic variables (CVs), based on the \chandra
observations of the LW.  However, the exact composition of the
discrete sources in the GRXE still remains unresolved \citep[see
also][]{Revnivtsev11}.  We explore the X-ray and optical properties of
the 10 periodic X-ray
sources in the LW and their implication
for evolutionary models of MCVs and their connection to thousands of
X-ray sources in the GCR.

\setcounter{footnote}{2}

\section{X-ray Observation and Timing Analysis} 

\subsection{\chandra Observation and Source Search} \label{s:obs}

The LW was observed for a total of 1 Ms exposure (100 ks in 2005
and 900 ks in 2008) with the \chandra ACIS-I instrument (H09b; R09).
Table~\ref{t:obs} lists the observational history of the field.  
The X-ray data were analyzed 
as a part of our ongoing \champlane survey
\citep{Grindlay05} and the analysis procedures are described in
detail in \citet[hereafter H05; see also H09b]{Hong05}. In summary, we stacked
all of the 13 separate pointings and searched for discrete sources 
in the 0.3--2.5, 2.5--8.0, and 0.3--8.0 keV band images,
using the {\it wavdetect} routine \citep{Freeman02}.  The source lists
of the three energy bands
were then cross checked to produce a list of 1397 unique discrete sources
based on the positional uncertainty of each source.  Carefully designed
aperture photometry (H05) that takes into account source crowding was applied
to each source in the 0.5--2.0, 2.0--8.0, and 0.5--8.0 keV bands.  The
complete source list and their photometries will be presented elsewhere
\citep[]{Hong11} along with comparison of the results from
two popular source detection algorithms, {\it wavdetect} and 
{\it wvdecomp}\footnote{By A.~Vikhlinin;
\url{http://hea-www.harvard.edu/RD/zhtools/}. See also M09.}.
In this paper, we 
use the sources discovered by the {\it wavdetect} routine. Since we can
identify periodic X-ray modulation only from relatively bright sources,
the two search
algorithms make no difference for the results of this paper.
The catalog of the 319
sources and their photometry results from the initial 100 ks
exposure can be found in H09b. 

\subsection{X-ray Timing Analysis} \label{s:analysis}

\begin{figure} \begin{center}
\includegraphics*[width=0.47\textwidth,trim = 7mm 0mm 0mm 0mm]{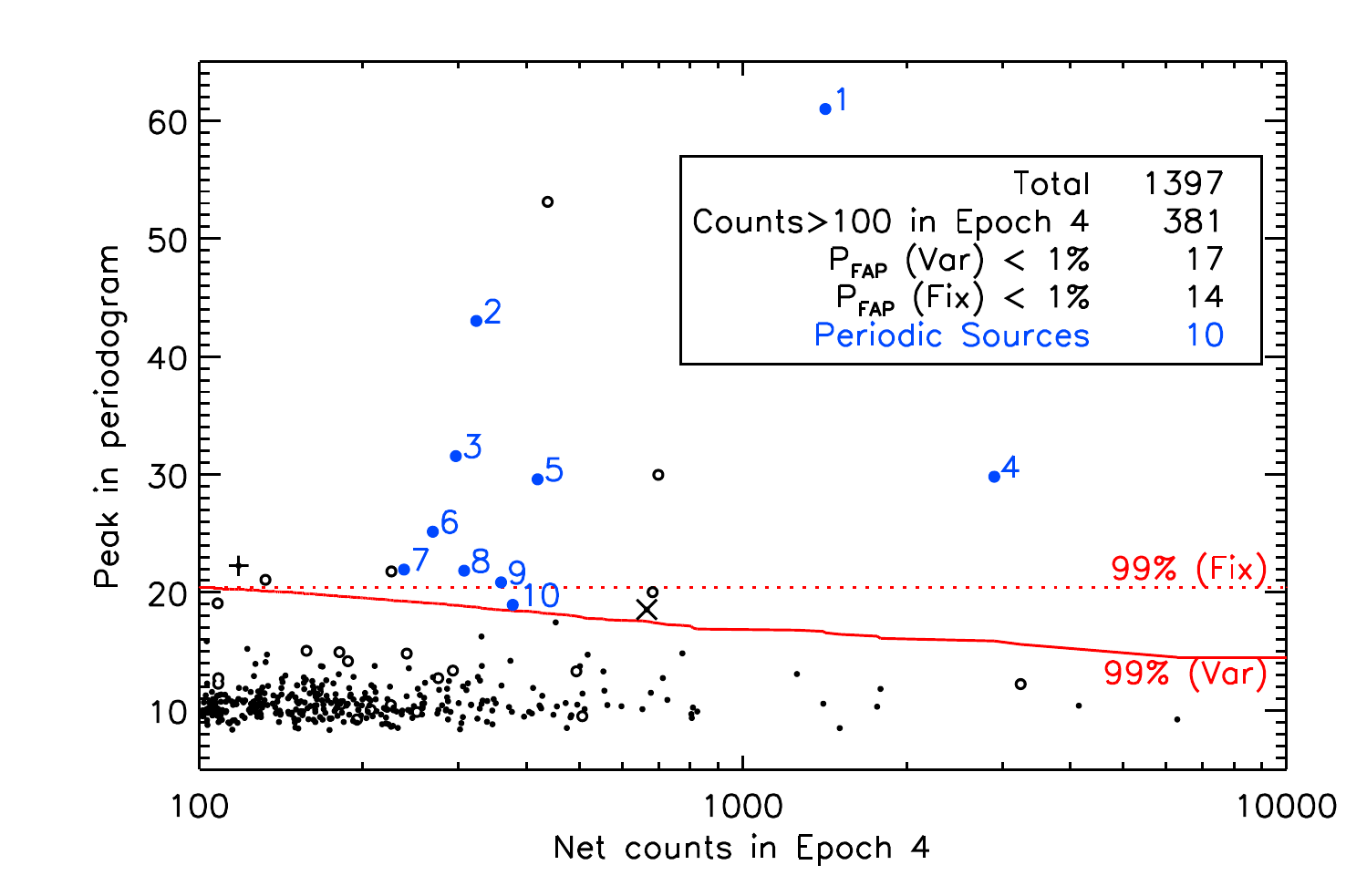}
\caption{The distribution of the background subtracted counts in Epoch 4 
of the sources and the peak values of the periodograms by the LS
method.  Of 17 sources with \fap (Var) $<$ 1\% (net count dependent
False Alarm Probability; see \S\ref{s:srcsel} for its definition) or
$P\Ss{conf}$ (Var) $>$ 99\% (red solid), we
identify 10
periodic sources (see \S\ref{s:xray}).  The periodicities of 5 sources (open
circles) are false, originating from either the dithering motion of the
instrument or non-periodic X-ray variabilities (e.g.~flares). 
We also exclude two other sources ('+' and 'x') because of the
marginal nature of their detections (see \S\ref{s:xray}).
}
\label{f:rundown}
\end{center}
\end{figure}

The 1 Ms \chandra exposure of the LW can be divided into
into four epochs so that the total span of each epoch does not exceed
a month, which allows period search in a single ephemeris
(Table~\ref{t:obs}).  Of four epochs, the last epoch provides
a long exposure (GTI: 801 ks) suitable for in-depth search of periodic
modulation.  For periodicity search, we selected the 381 sources with
background-subtracted net counts greater than 100 in the 0.3--8.0 keV
band in
Epoch 4 (Fig~\ref{f:rundown}).  

Photon arrival times of each source were Barycenter
corrected to Barycentric Dynamical Time (TDB) by the {\tt axbary} routine
in the CIAO tool (ver 3.4).  Then, we generated a background-subtracted
light curve of each source in 12.8 s bins, and we
applied a Lomb-Scargle (LS) routine \citep{Scargle82} to the light curve
to search for periodic X-ray modulation.  We subtracted the background
counts in the source aperture region using the events, with a proper
scale, that fell in the background annulus region around the source,
which excludes the source regions of neighbors (H05).  The 12.8 s bin
interval was chosen for efficient executions of the LS routine over a
long exposure, and it only suppresses short periods below $\sim$ 100 sec (see
\S\ref{s:nsmcvs} and Fig.~\ref{f:pbias}), where there is already a
concern due to the CCD readout time ($\sim$ 3.2 s).  Randomization of
photon arrival times to compensate for fixed CCD readout cycle makes practically
no difference for the search results due to the longer time bins used for
the light curves and
the long exposure consisting of multiple pointings, where 
multiple phases of CCD readout cycle 
produces an effect similar to randomizing arrival times to some degree.

The search periods were selected successively,
starting from the total duration ($T$) down to $10^{-4} T$ by a decrement
of $\Delta P = P^2/ (2 T s_f)$, where we introduced an oversampling
factor, $s_f$ ($s_f$=1 means no oversampling), in order to sample the
periodograms relatively smoothly over the entire period search range.
We change $s_f$ logarithmically
from 1.0 at the shortest period ($10^{-4} T$) to 4.0 at the longest
($T$), and a relative increase in the number of search periods due to
the introduction of $s_f$ is about 30\%.  In Epoch 4
where the total duration spans about 17.7 days with a combined total exposure
of 832 ks, we searched 26499 periods from about 153 s to 17.7 days.
In this analysis, we further limit the search range between 153 s
and 10 hr, which contains 19911 independent periods out of 26247
trial periods. This
range covers most of the spin and orbital periods of MCVs. We found
that apparent periodicities of $>$ 10 hrs 
indicated by the periodograms
are false alarms, usually caused by short or long term X-ray variabilities
(e.g.~flares).  For selected periodic sources (\S2.3), we also extended
the period search range down to 20 s in attempt to find any secondary
modulation at periods shorter than 153 s, although the bin size (12.8
s) of the light curves may suppress detection of short periods below
$\sim$ 100 s (see Fig.~\ref{f:pbias}).

For candidate periodic sources, we also
applied an Epoch Folding (EF) method \citep{Leahy83} to refine the
periods and compare the significance of the primary periods with
their harmonics.  For the EF method, we generated a background-subtracted
folded light curve in 15 bins for a given search period and calculated the
$\chi^2$ value of the folded light curve with respect to the assumed constant count
rate of no periodicity.  For each source, we applied the EF method at 1000
periods within 2 $\sigma$ of the primary period found by the LS routine,
and the same around the half, 2nd and 3rd harmonics of the primary period.

\subsection{Periodic X-ray Source Selection} \label{s:srcsel}

In order to select valid periodic sources, we have estimated a 
False Alarm Probability (\fap) 
or a confidence level ($P\Ss{conf} \equiv 1 - $\fap).  For a given 
power density ($X$) in the LS routine, \fap$=1-(1-e^{-X})^{N_S N_P}$,
where $N_P$ is the number of independent trial periods (19911) and
$N_S$ is the number of the sources.  To validate \fap estimated from
the above trial statistics, we have also performed $\sim$ 100,000 simulations
by randomizing photon arrival times.  These simulations allow reliable
estimates of \fap down to $\sim$ 0.1\% for $N_S \lesssim 10$ or $\sim$
1\% for $N_S \lesssim 100$, etc., where the simulation results match
with those from the trial statistics. For a larger value of 
$N_S$ or a smaller value of \fap, more
simulations are required.  In the following, we calculate \fap according to
the trial statistics.

The search is usually conducted for the brightest sources first, since
they have a better chance of periodicity detection.  Depending on the
search limit of net count ($N\Ss{net}$), which determines $N_S$,
\fap changes. Although the LS routine is known to be more sensitive
to low count sources than a few other conventional Fourier
analyses \citep[e.g.~see Chapter 6.1 in][]{Bretthorst88}, 
its performance eventually drops out as the net count
decreases to zero. Extending the search limit by including low count sources
with no chance of periodicity detection can arbitrarily increase \fap,
which can make valid periodic modulation appear insignificant.  On the
other hand, since search performance depends on various factors
(e.g.~background counts) besides net counts, it is desirable to
extend the search to a relatively low net count to be conservative. 
Therefore, we set the search limit
at a relatively low value ($N\Ss{net}$ = 100), and we use
two types of the \fap estimates: \fap (Var), where
$N_S=N_S$($\ge$$N\Ss{net}$), and \fap (Fix), where
$N_S=N_S$($N\Ss{net}$ $\ge$100) = 381.
The former provides an initial cut for candidate periodic
sources, and the latter provides a list of sources with a higher confidence.


Modulation amplitude is also a good metric for significance of periodicity.
In order to describe diverse pulse profiles, we use the \rms
amplitude ($A\Ss{RMS}$) given by
\begin{eqnarray}
	A\Ss{RMS} &=& \left(\frac{Z^2_1}{N}\right)^{1/2} \frac{N}{N-B}, \\
	Z^2_1 &=& \frac{2}{N} \left\{\left[\Sigma_j \cos\phi(t_j)\right]^2 
		+ \left[\Sigma_j\sin^2\phi(t_j)\right]^2\right\},
\end{eqnarray}
where $N$ and $B$ are the total and background number of events 
in the source aperture region respectively, and $t_j$ is
the arrival time, and $\phi(t)$ is the phase at time $t$ that would be
expected for a given period \cite[hereafter M03b and the references therein]{Muno03b}.
For easy comparison with the literature, we also define
another modulation amplitude as 
	$A_M = 1 - R\Ss{min}/R\Ss{max}$, where
$R\Ss{min}$ ($R\Ss{max}$) is the minimum (maximum) count rate of the
folded light curve. 
Finally, for synthetic light
curve simulations, we use sinusoidal variations, where the light curve is described
as $1 + A_0 \sin \phi(t)$, where $0 <  A_0 \le 1$ and $\phi(t)$ is the phase
of photon arrival time $t$. 
For sinusoidal modulations,
$A_0$ = $\sqrt{2} A\Ss{RMS}$,
$A_M$ = 2 $A_0/(1+A_0)$, and $A\Ss{RMS} < A_0 \le A_M$.
Since both $R\Ss{min}$ and $R\Ss{max}$
depend on the bin size of folded light curves, we calculate $A_M$
from $A\Ss{RMS}$ using the above relations.

We also use simulations to calculate a detection probability of
periodicity ($P\Ss{det}$). For each candidate periodic source,
we generate 1000 synthetic light curves to estimate
$P\Ss{det}$ within about 1\% accuracy.  Each synthetic light
curve is consistent with a sinusoidal variation of the measured period
and modulation amplitude of the source, while accounting for the
GTIs and Barycentric shifts of the CCD readout time of the real data.
Then we perform the LS search algorithm to see how often we detect the
same periodicity with
$P\Ss{FAP}$ (Fix) $\le 1\%$.
These simulations enable completeness correction
of periodicity detection (see also \S\ref{s:complete}) and provide
another validity measure of the detection in addition to \fap.

\begin{table*}
\small
\caption{Periodic sources in the LW}
\begin{tabular*}{\textwidth}{c@{\extracolsep{\fill}}cD{(}{(}{5.3}D{(}{(}{5.3}rrrccl}
\hline\hline
(1) & (2)              & \mbox{(3)}                           & \mbox{(4)}                 & (5)         & (6)         & (7)         & (8)       & (9)       & (10)         \\
ID  &  Source Name     & \multicolumn{1}{c}{Counts (Epoch 4)} & \multicolumn{1}{c}{Period} & \fap (Fix)  & $A$\Ss{RMS} & $P$\Ss{det} & Offset    & Signi.    & Notes        \\
LWP & CXOPS            & \multicolumn{1}{c}{(0.3--8 keV)}     & \mbox{(s)}                 & ($10^{-n}$) & (\%)        & (\%)        & (\arcmin) & Harmonics &              \\
\hline
1   & J175151.2-293310 & 1418(41)                             & 10342(5)                   & 19.6        & 26(3)       & 99.8        & 5.6       &           &              \\
2   & J175123.5-293755 & 323(21)                              & 5131(5)                    & 11.8        & 59(5)       & 99.9        & 2.6       &           &              \\
3   & J175129.1-292924 & 296(22)                              & 7448(10)                   & 6.8         & 65(6)       & 99.1        & 6.1       &           &              \\
4   & J175131.6-292956 & 2899(56)                             & 8536(14)                   & 6.1         & 14(2)       & 83.3        & 5.6       &           & BB:4         \\
5   & J175133.9-292754 & 419(29)                              & 6342(7)                    & 6.0         & 43(4)       & 67.5        & 7.7       &           &              \\
6   & J175118.7-293811 & 269(19)                              & 4729(2)                    & 4.0         & 50(5)       & 80.1        & 3.3       &           &              \\
7   & J175122.7-293436 & \sS{\dagger}119(13)                  & 12076(91)                  & 2.6         & 60(8)       & 25.7        & 1.5       & 2         & Edge: 352 ks \\
8   & J175147.4-294215 & 307(28)                              & 4890(6)                    & 2.4         & 40(4)       & 9.6         & 7.9       &           &              \\
9   & J175133.6-293313 & 359(21)                              & 6597(10)                   & 2.1         & 39(5)       & 66.1        & 2.6       & 3         & PO:0.91      \\
10  & J175119.4-293659 & 377(22)                              & \sS{\ddagger}5262(2)       & 1.4         & 30(5)       & 13.9        & 2.4       & 2         & 2nd harmonic \\
\hline
\end{tabular*}

\label{t:detection}
(1) An abbreviated X-ray source ID.
(2) The \chandra source name.
(3) The background subtracted net counts in the 0.3--8 keV band in
Epoch 4 (832 ks exposure).
\sS{\dagger}The counts for LWP~7 is from the last 352 ks GTI, which is free of
near CCD edge events (see \S\ref{s:lwp7}).
See Table~\ref{t:spec} for the total net counts of the 1 Ms exposure.
(4) The modulation periods are refined by 
the EF routine around the significant periods found by the LS
routine.
\sS{\ddagger}For LWP 10, the second harmonic (5262 s) 
is considered the real period (see \S\ref{s:lwp10}).
(5) The False Alarm Probability for $N_S$=381 (see \S\ref{s:srcsel}). 
(6) The \rms modulation amplitude based on Eq.~1. 
(7) The detection probability of periodicity based on 
simulations using 1000 synthetic light curves for each source.
(8) The offset from the aimpoint of Obs.~ID 5934 (R.~A.: 17h 51m
28.50s, DEC.: -29\Deg 35\arcmin 16.80\arcsec), 
(9) The most significant harmonic according to the EF method if the primary
period is not. In LWP~10, the second harmonic is significantly more
prominent than the primary primary (see \S\ref{s:lwp10}), whereas in
the other two sources, the differences are marginal.
(10) 
BB: the number of independent Bayesian Blocks in the light curve if not one (see \S\ref{LWP4}).
Edge: the source falls near a CCD edge in some pointings and GTI free
of such pointings is shown.  
PO: the 95\% PSF overlaps with neighbors. The number indicates
the fractional radius of the non-overlapping region (H05).
\end{table*}


Finally, we also performed the Bayesian Block (BB) search for long term
variability \citep{Scargle98}.  In order to apply the BB search to
the data set of a long duration (3 yr) with long exposure gaps ($>$1 yr), we
eliminate exposure gaps longer than 20 ks, and generate a semi-contiguous
series of photon arrival times.  

The simulations for \fap and \detp along with period search
require significant computational power.  In order to handle
the computational burden efficiently at low cost, we have utilized a
GPU-based desktop supercomputer equipped with a C1060
Tesla GPU (240 cores) from Nvida\footnote{\url{http://www.nvidia.com}}.
A simple code conversion using an
IDL\footnote{\url{http://www.ittvis.com/idl}} GPU library ({\it
gpulib}\footnote{\url{http://txcorp.com/products/GPULib}}) has boosted
the speed of some routines by a factor of 10 by comparison to the
regular IDL routines\footnote{The complete code optimization is expected to
boost the speed by another factor of 10 according to the examples
given by {\it gpulib}.}.

\section{X-ray Analysis Results} \label{s:xray}

Fig.~\ref{f:rundown} illustrates the search results and the periodic
source count.  For initial screening, we select 17 sources that exhibit
a periodicity with \fap (Var) less than 1\%.  We exclude five sources
with apparent periodicity due to either the dithering motion of the
telescope (707, 1000 s and their immediate (sub-) harmonics), or 
non-periodic variabilities (e.g.~flares).  In addition, we exclude
J175055.5-292948 (period: 5331 s) since its $P\Ss{det}$ is only 0.1\% ('+' in
Fig.~\ref{f:rundown}).  For sources with \fap (Fix) $\ge$ 1\%, we only
consider them periodic if they show another significant indicator
of periodicity (e.g.~the results of the EF method, see
\S\ref{s:lwp10}), and so J175103.9-293430 (period: 3821 s) is also
excluded ('x' in the figure).  The final list of the periodic sources in the LW contains 10
sources (9 with \fap (Fix) $<$ 1\%).  The spatial distribution of the
10 periodic sources does not show any obvious sign of clustering or pattern.

Table~\ref{t:detection} lists the
basic results of period search and photometry of the 10 periodic sources in the LW.  For easy reference, in
addition to the X-ray source name (starting with CXOPS, see H05), we
assign an abbreviated version of the name starting with prefix 'LWP'
indicating 'Limiting Window Periodic sources'. The number IDs in the
short name are assigned in the reverse order of the periodogram value
at the observed periods (i.e. the likelihood of true periodicity).
Note B09 list the sources with prefix 'LW'.

\begin{figure*} \begin{center}
\includegraphics*[width=1\textwidth]{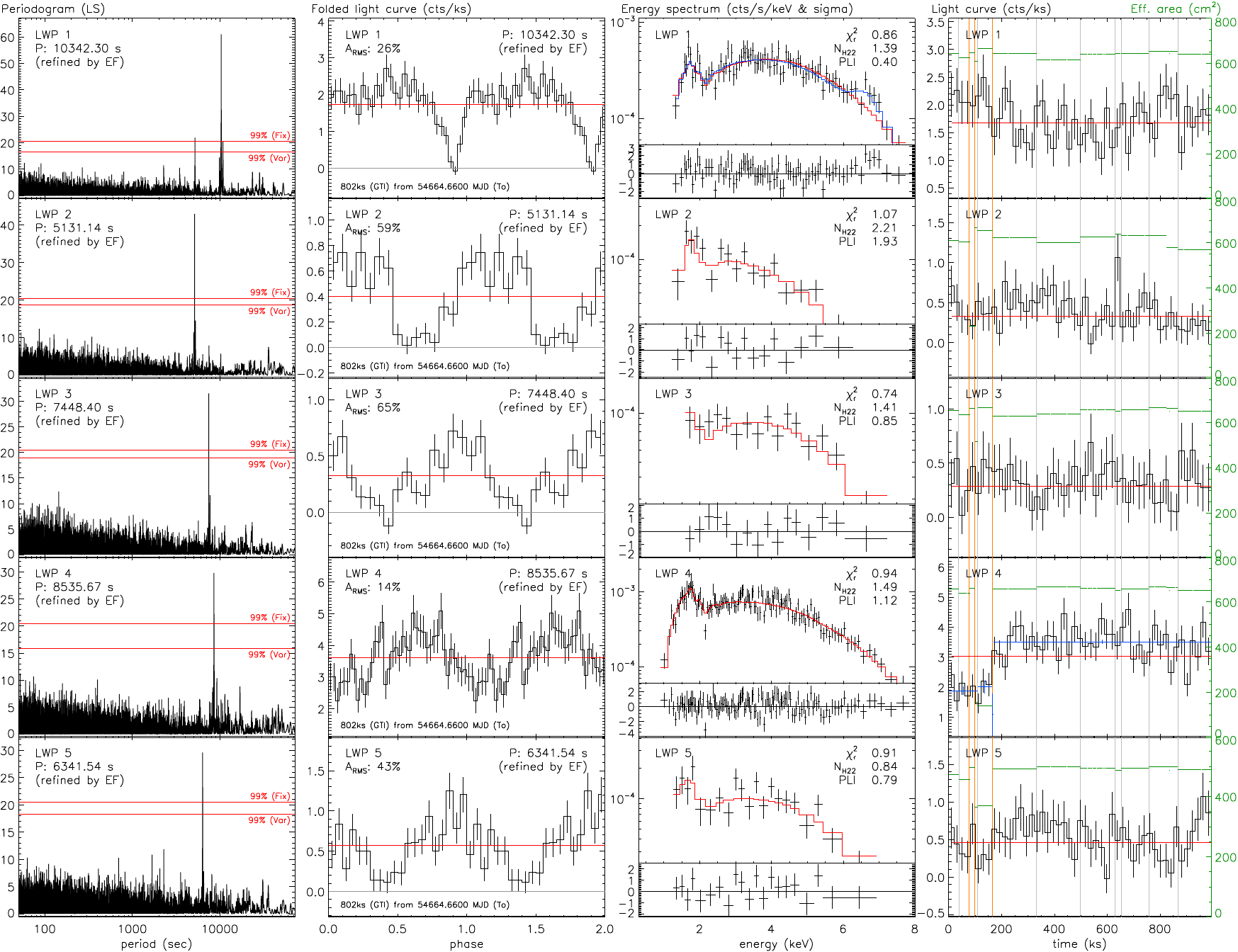}
\caption{Periodograms, folded light curves, energy spectra and
compressed light curves of the periodic X-ray sources (LWP 1--5)
in the LW.  The periodograms are based on the LS method, and
the 99\% confidence levels for $N_S=381$ (Fix) and $N_S=N_S(>N\Ss{net})$ (Var) 
are shown in (red) horizontal lines (see \S\ref{s:srcsel}).
The folded light curves are drawn with T\Ss{0} = MJD
54664.6600, and the (red) horizontal
lines show the average rates. In the energy spectra, the (red) lines
show the results of simple power law model fits. In the case of LWP 1,
a model fit using a power law plus an iron line is shown in blue. In the
compressed light curves, the (red) horizontal lines represent the average
rate, the (green) steps show the effective area of each pointing, and the
vertical lines indicate the epoch boundaries (yellow)
and exposure gaps (grey, $>20$ ks).
For LWP 4, the Bayesian Blocks are shown in blue lines.  }
\label{basic1}
\end{center}
\end{figure*}

\begin{figure*}[t] \begin{center}
\includegraphics*[width=1\textwidth]{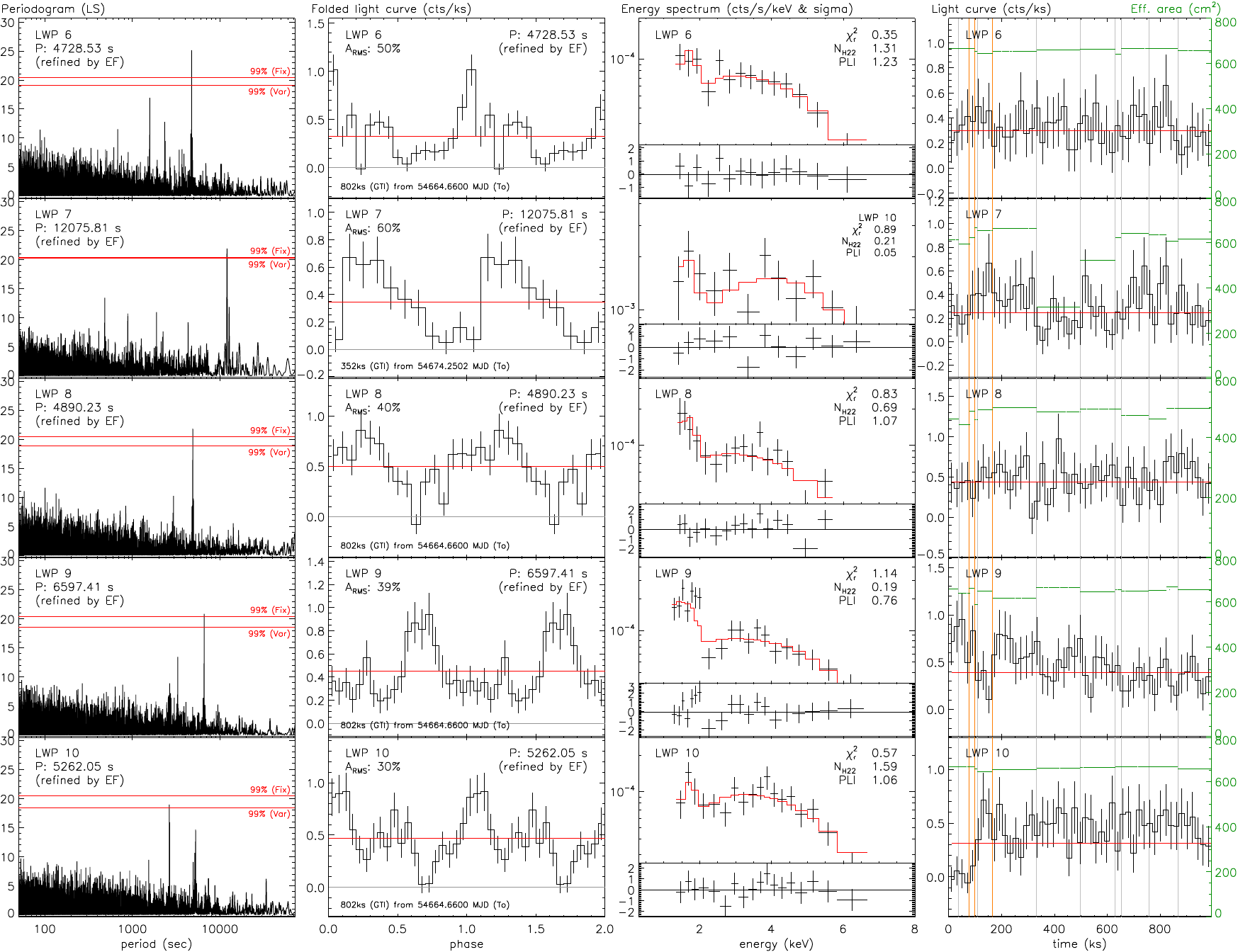}
\caption{Same as Fig.~\ref{basic1} for 
LWP 6--10. The folded light curve of LWP~7 is drawn with T\Ss{0} =
MJD 54674.2502.
}
\label{basic2}
\end{center}
\end{figure*}

Figs.~\ref{basic1} and \ref{basic2} show the periodograms, the folded
light curves, the X-ray spectra with model fits, and the compressed
light curves without long exposure gaps of the 10 periodic sources. The
periodograms are based on the LS method, which was applied to the binned
light curves of Epoch 4.  The folded light curves are also generated from
the data obtained in Epoch 4 and are very similar to the ones from the
full 1 Ms exposure (not shown).  The periodogram and the folded light
curve of LWP 7 are from the last 352 ks GTI in 8.1d when the source was
observed sufficiently distant of any CCD edges.

The X-ray spectrum of each source is generated using all of the available
data, and each spectral bin contains more than 20 background-subtracted
net counts.  The red solid lines show the results using a simple power
model fit, and for LWP~1, a spectral fit using a power law plus an iron
line at 6.7 keV is also shown (blue).  In the compressed light curves,
the long exposure gaps ($>$ 20 ks), are marked by the vertical lines,
and the yellow lines indicate the boundaries of four epochs.  The red
horizontal lines show the average count rate and the green lines show
the effective area, which varies from pointing to pointing. The blue
solid lines for LWP~4 show the count rates of the BBs.

Table~\ref{t:spec} summarizes the X-ray spectral properties including
estimates of the unabsorbed X-ray fluxes, based on
spectral model fits. Note the net counts in Table~\ref{t:spec} are from the
full 1 Ms exposure, whereas the net counts in Table~\ref{t:detection} are
from Epoch 4 except for LWP~7, which shows the net count from the last 352
ks GTI.  We fit the spectra of all the sources using a simple power law
model, and for LWP~1, we also tried a power law plus an iron line (see \S\ref{LWP1}).


\section{Optical Observation and Properties} \label{s:optprop}

We observed the LW field with the \HST and Magellan telescope.  Five of
the 10 sources were in the \HST/ACS fields (B09), and three of them were
reported as possible candidates for accreting binaries or CVs.
Table~\ref{t:opt} summarizes the combined optical and X-ray properties
of the candidate optical counterparts of the \chandra sources.  
For easy reference, the short names used by B09 are noted for the
three sources in
Table~\ref{t:opt} (e.g. LWP~2 = LW~25).

\subsection{\HST/ACS Data and Analysis}

We observed with \HST the inner area of the ACIS field with a 2 \x 2
mosaic of the Wide Field Camera (WFC) on the ACS on 2005 August 19. A
single WFC pointing images a 3.4\arcmin \x 3.4\arcmin\ field with $\sim$
0.05\arcsec\ pixels using two CCD detectors separated by a 2.5\arcsec\
gap.  Exposures were taken through the F435W
($B\Ss{435}$), F625W ($R\Ss{625}$, similar to Sloan $r$), and F658N
(H$\alpha$) filters. Each tile of the mosaic was observed with the same
exposure sequence 4 \x 492 s in F435W, 168 s + 2 \x 167 s in F625W, and
4 \x 496 s + 4 \x 492 s in F658N.  No dithering was applied to fill in
the WFC chip gap.  Photometry is performed using a stellar-photometry
package, DOLPHOT, a modified version of the HSTphot package
to do photometry on HST/WFPC2 images \citep{Dolphin00}.
See B09 for the details.

\subsection{Magellan/IMACS Data and Analysis}

On 2007 May 8, we observed the LW field and two other Window fields
(Stanek's and Baade's) with the Inamori Magellan Areal Camera and
Spectrograph (IMACS) on the 6.5 m Magellan (Baade) telescope at Las
Campanas, Chile. With seeing $\sim 0.8-1.2\arcsec$ FWHM,
we obtained a dithered set of 5 pointings in the f/4
configuration (15.4\arcmin\ field, 0.11\arcsec/pixel) to cover an $18\arcmin
\times 18\arcmin$ region of the LW. This provided a total exposure time of
600, 300, 180, \& 180 s in Bessell-$B$, $V$, $R$, \& CTIO-$I$
filters over the \chandra field respectively.

We processed the images using standard IRAF tasks, and calibrated the
astrometry using the 2MASS catalog as a reference. The astrometric
residuals on each CCD frame were $\sim$ 0.2\arcsec. We reprojected and
stacked the images using the
SWARP\footnote{\url{http://www.astromatic.net/software/swarp}}
utility. All frames were normalized to ADU/second units and combined
using weight-maps constructed from flat-fields and bad pixel masks. The
initial source search and photometry were performed on the stacked images
using DAOPHOT.  See also H09a.

\subsection{Optical Matches}

Both the \HST and IMACS source lists are boresighted to the \chandra
sources as described in \citet{Zhao05}: the boresight correction is less
than 0.1\arcsec\ in both R.~A. and Declination.  In search of optical
counterparts and their optical properties, for the five sources in
the \HST/ACS fields, we use the \HST observations. For the other five,
we use the Magellan/IMACS observations, where three sources show one
or two candidate counterparts. Considering the high stellar density in
the region, there is no guarantee of these Magellan/IMACS sources being
the true counterparts (e.g.~LWP~9 has six candidate counterparts seen
in the \HST/ACS field, but only one in the Magellan/IMACS image).

The X-ray-to-optical flux ratios, Log$(F_X/F_R)$, in Table~\ref{t:opt}
are calculated for the 0.3-8 keV band vs.~the $R$ magnitude.  Both the
observed and unabsorbed flux ratios are given.  For unabsorbed flux
ratios, the $R$ magnitude is de-reddened based on the $A\Ss{F625W}$ map
given by \citet{Revnivtsev10}.  For sources with multiple candidate
counterparts, a range of the $R$ magnitude and the flux ratio is given,
covering all the candidate counterparts.  In the case of LWP~2 (= LW~25)
and LWP~6 (= LW~8),
shown is the $R$ magnitude of the most likely
candidate with an unusually blue color (B09).  In the case of LWP~8,
only one of the candidate counterparts has a measurable $R$ magnitude.
For sources with no detectable candidate counterparts, a lower
limit for the $R$ magnitude is given, based on the minimum value of 
optical neighbors within a 30\arcsec\ radius of the X-ray source position.

\begin{table*}[t]
\small
\caption{X-ray spectral properties of the periodic sources in the LW 
based on simple power model fits}
\begin{tabular*}{\textwidth}{c@{\extracolsep{\fill}}D{(}{(}{4.3}D{.}{.}{2.5}D{.}{.}{2.4}D{.}{.}{3.4}D{.}{.}{2.4}D{/}{/}{3.3}D{(}{(}{3.4}D{(}{(}{3.4}D{(}{(}{3.4}c}
\hline\hline
\multicolumn{1}{c}{(1)} & \multicolumn{1}{c}{(2)}            & \multicolumn{1}{c}{(3)}      & \multicolumn{1}{c}{(4)}  & \multicolumn{1}{c}{(5)}      & \multicolumn{1}{c}{(6)}   & \multicolumn{1}{c}{(7)}                &                                                           & (8)                      &                                & (9)                                         \\
Source                  &                                    &                              &                          &                              &                           &                                        & \multicolumn{3}{c}{Unabsorbed Flux} &                                                                          \multicolumn{1}{c}{Luminosity}              \\
\cline{8-10} ID         & \multicolumn{1}{c}{Counts (Total)} & \multicolumn{1}{c}{$E_{50}$} & \multicolumn{1}{c}{\nHt} & \multicolumn{1}{c}{$\Gamma$} & \multicolumn{1}{c}{EW}    & \multicolumn{1}{c}{$\chi^2_{\nu}$/DoF} & \multicolumn{1}{c}{0.5--2}                                & \multicolumn{1}{c}{2--8} & \multicolumn{1}{c}{0.3--8 keV} & \multicolumn{1}{c}{0.3--8 keV}              \\
LWP                     & \multicolumn{1}{c}{(0.3--8 keV)}   & \multicolumn{1}{c}{(keV)}    &                          &                              & \multicolumn{1}{c}{(keV)} &                                        & \multicolumn{3}{c}{(10\sS{-15} erg cm\sS{-2} s\sS{-1})} &                                                      \multicolumn{1}{c}{(10\sS{x} erg s\sS{-1})} \\
\hline
1                       & 1775(45)                           & 3.86(6)                      & 1.4(2)                   & 0.4(1)                       &                           & 0.86/74                                & 6.3(5)                                                    & 56(2)                    & 63(2)                          & 30.9 -- 32.7                                \\
                        &                                    &                              & 1.7(1)                   & 0.63(2)                      & 0.5(1)                    & 0.74/72                                & 8.3(7)                                                    & 55(1)                    & 64(2)                          & 30.9 -- 32.7                                \\
2                       & 393(23)                            & 2.9(1)                       & 2.2(5)                   & 1.9(3)                       &                           & 1.07/14                                & 9(1)                                                      & 8.9(6)                   & 21(1)                          & 30.4 -- 32.2                                \\
3                       & 361(24)                            & 3.7(2)                       & 1.4(9)                   & 0.8(4)                       &                           & 0.74/13                                & 1.4(4)                                                    & 10.2(7)                  & 12.0(8)                        & 30.2 -- 32.0                                \\
4                       & 3217(59)                           & 3.36(4)                      & 1.5(1)                   & 1.12(8)                      &                           & 0.94/123                               & 24(1)                                                     & 79(2)                    & 107(2)                         & 31.1 -- 32.9                                \\
5                       & 473(30)                            & 3.4(2)                       & 0.8(5)                   & 0.8(3)                       &                           & 0.91/18                                & 2.5(4)                                                    & 12.7(9)                  & 16(1)                          & 30.3 -- 32.1                                \\
6                       & 329(21)                            & 3.2(1)                       & 1.3(6)                   & 1.2(4)                       &                           & 0.35/11                                & 2.6(4)                                                    & 7.0(5)                   & 10.0(6)                        & 30.1 -- 31.9                                \\
7                       & 295(20)                            & 3.8(1)                       & 0.0(3)                   & 0.0(1)                       &                           & 1.00/10                                & 0.32(6)                                                   & 7.8(6)                   & 7.6(5)                         & 30.0 -- 31.8                                \\
8                       & 361(30)                            & 3.0(2)                       & 0.7(8)                   & 1.1(5)                       &                           & 0.83/13                                & 1.9(4)                                                    & 8.1(7)                   & 10.3(9)                        & 30.1 -- 31.9                                \\
9                       & 457(24)                            & 2.9(1)                       & 0.2(3)                   & 0.8(3)                       &                           & 1.14/17                                & 1.3(1)                                                    & 8.0(5)                   & 9.3(5)                         & 30.0 -- 31.9                                \\
10                      & 415(23)                            & 3.58(9)                      & 1.6(5)                   & 1.1(3)                       &                           & 0.57/15                                & 3.1(4)                                                    & 9.7(6)                   & 13.1(7)                        & 30.2 -- 32.0                                \\
\hline
\end{tabular*}

\label{t:spec}
(1) An abbreviated X-ray source ID.
(2) The background-subtracted net counts in the 0.3--8 keV band from
the full 1 Ms exposure.
(3) The median energy of the photons in the 0.3--8 keV band.
(4) \& (5) An estimate of the power law index ($\Gamma$) and the absorption
(\nHt) from spectral model fits, 
assuming a simple power law model for the X-ray spectrum in the
0.3--8 keV band. 
(6) The estimated equivalent
width of the iron line from a spectral fit of a power law model with an
iron line at 6.7 keV.
(7) Reduced $\chi^2$ and Degree of Freedom (DoF) of spectral model fits.
(8) An estimate of the unabsorbed flux.
(9) The luminosity range in the 0.3--8 keV band for source distance
at 1--8 kpc.
\end{table*}

\begin{table*}
\small
\caption{Candidate optical counterparts of the periodic sources in the LW}
\begin{tabular*}{\textwidth}{c@{\extracolsep{\fill}}ccccccccc}
\hline\hline
(1)                                   & (2)          & (3) & \multicolumn{2}{c}{(4)} &                           \multicolumn{2}{c}{(5)} &                       \multicolumn{2}{c}{(6)} &                      (7)           \\
Source                                & R            & B-R & \multicolumn{2}{c}{$F_X$ (0.3--8 keV)} &            \multicolumn{2}{c}{Log$(F_X/F_R)$} &            \multicolumn{2}{c}{No.~of Candidates} &        $A$\Ss{F625W} \\
\cline{4-5}\cline{6-7}\cline{8-9} LWP & (mag)        &     & (observed)                               & (unabsorbed) & (observed)                           & (unabsorbed) & \HST                                    & Magellan & (offset)      \\
\hline
1                                     & $>$22.3      &     & 51(1)                                    & 64(2)        & $>$1.4                               & $>$-0.1      &                                         &          & 3.9 (1.5$'$)  \\
2                                     & 22.3 .. 24.4 &     & 8.0(5)                                   & 21(1)        & 0.6 .. 1.4                           & -0.4 .. 0.4  & 2 (LW25)                                &          & 3.6 (0.1$'$)  \\
3                                     & $>$22.2      &     & 9.2(6)                                   & 12.0(8)      & $>$0.6                               & $>$-1.0      &                                         &          & 4.2 (1.9$'$)  \\
4                                     & 23.1(4)      &     & 75(1)                                    & 107(2)       & 1.9(2)                               & 0.4(2)       &                                         & 1        & 4.1 (1.4$'$)  \\
5                                     & 20.5(2)*     &     & 13.0(8)                                  & 16(1)        & 0.08(8)*                             & -1.47(8)*    &                                         & 1        & 4.1 (3.4$'$)  \\
6                                     & 22.7(1)      & 2.9 & 6.8(4)                                   & 10.0(6)      & 0.68(5)                              & -0.72(5)     & 5 (LW8)                                 & 1        & 3.9 (0.3$'$)  \\
7                                     & 18.8         &     & 7.6(5)                                   & 7.6(5)       & -0.83(3)                             & -2.47(3)     & 1                                       & 1        & 4.1 (0.5$'$)  \\
8                                     & 21.4(2)      &     & 8.1(7)                                   & 10.3(9)      & 0.23(9)                              & -1.02(9)     &                                         & 2        & 3.4 (3.5$'$)  \\
9                                     & 22.4 .. 25.7 &     & 8.6(5)                                   & 9.3(5)       & 0.7 .. 2.0                           & -0.7 .. 0.6  & 6 (LW19)                                & 1        & 3.6 (0.2$'$)  \\
10                                    & 22.6 .. 24.6 &     & 9.3(5)                                   & 13.1(7)      & 0.8 .. 1.6                           & -0.6 .. 0.2  & 4                                       &          & 3.8 (0.1$'$)  \\
\hline
\end{tabular*}

\label{t:opt}
(1) An abbreviated X-ray source ID.
(2) $R$ magnitude, based on the \HST observation if available: otherwise,
based on the Magellan Images. For sources with multiple counterparts,
the range of those counterparts is given.
For sources with no valid counterparts,
a lower limit is given, based on the minimum magnitude of 
neighboring optical sources within a 30\arcsec\ radius from the X-ray
source position. '*' uses the $I$ magnitude instead of $R$.
(3) $B$ (F435W) $-$ $R$ (F625W) from the \HST observation.
(4) The X-ray flux in 10\sS{-15} erg cm\sS{-2} s\sS{-1}, 
from column (7) in Table~\ref{t:spec}.
(5) Log$(F_X/F_R)$ = Log$(F_X)+R/2.5+5.76$. For unabsorbed flux ratios,
the $R$ magnitudes are de-reddened by 
$A\Ss{F625W}$.
(6) The number of candidate counterparts found in
the Advanced Camera for Surveys (ACS) images of the \HST
and the Magellan MOSAIC Images.
Every source in the ACS FoV has more than one counterpart and
for sources reported by B09, their source IDs are given.
(7) $A\Ss{F625W}$ from \citet{Revnivtsev10} and the offset of the source from
where $A\Ss{F625W}$ is sampled.
\end{table*}

The logarithmic flux ratios or their limits of these periodic sources
are relatively high: $> -1$ for 7 sources and $> -2.5$ for the rest.
The intrinsic X-ray-to-optical flux ratios of accreting binaries or
AGN are usually significantly higher than those of coronal sources (see
\S4.1.2 in B09 and the references therein).  CVs have the intrinsic flux
ratio between $-2.5$ and +0.5 in the above energy bands, although a few
outliers of active binaries or dMe stars have the flux ratios as high
as $-1$ or $-0.5$. Therefore, the flux ratio results in
Table~\ref{t:opt} are consistent with those of CVs.

Note that, except for a few cases with an outstanding blue counterpart,
most of the flux ratio values in
Table~\ref{t:opt} are in fact likely lower limits, considering
the high stellar density in the region and the possibility of the true
counterpart being fainter and undetected.  LWP~5 and 8 have
additional uncertainties in their estimates due to the variation in
the interstellar
absorption across the region: for these sources, the $A\Ss{F625W}$ values
were sampled about 3 or 4\arcmin\ away from the sources.  For LWP~7 and 9,
the absorption in the X-ray spectra is estimated less than what is
expected in the field, based on $A\Ss{F625W}$. For LWP~3, 5 and 8, two
estimates are consistent, and for the rest, the former is larger than
the latter.

\section{Source Properties} \label{s:each}

In this section, we discuss some of the unique properties or
analysis caveats of each source. 

\subsection{LWP~1: CXOPS J175151.2-293310}\label{LWP1}

The periodogram reveals significant periodic modulation at the primary
period and its half.  The folded light curve shows a clear eclipse, 
but due to the lack of other features in the folded light curve other than the
eclipse, the system can be either a polar or an IP.
The long period suggests the modulation is likely due to the orbital
motion (e.g. only two IPs with $\gtrsim$
10\sS{4} s spin period in Fig.~\ref{pdist}a, see \S6.1 for the details).
If so, the mass of the companion is estimated to be $\sim$ 0.42 $\Ms$
according to Eq.~2.89 in \citet{Warner95}.
In addition, from the eclipse duration ${\Delta}\phi \sim 0.09$, the
mass ratio $q \gtrsim 0.29$ and the inclination angle $i$ is somewhere
between 60\Deg and 74\Deg for white dwarf (WD) mass
$M_1\sim 0.1-1.4\Ms$ \citep[Eq.~2.92 \& 2.93 in][]{Warner95}.

The X-ray spectrum exhibits a feature consistent with an iron line at 6.7
keV (and perhaps 6.4 keV as well). The spectral fit with a power law plus a 6.7 keV line
reduces the $\chi^2_{\nu}$ by 14\%, compared to a simple power law model
(Table~\ref{t:spec}).  In order to estimate the
significance of the line \citep[see][]{Protassov02}, we generated 1000
synthetic spectra, each of which is consistent with a simple power law
model ($\Gamma=0.4$ and \nHt = 1.4). We then fit each spectrum with a power
law plus the iron line, and
count how many cases reduce the $\chi^2_{\nu}$ by more than 14\%.
The results were none, indicating the significance
of the line is 99.9\% or higher.  
The hard X-ray spectrum with an iron emission line suggests the source is
likely an IP
(see more about the source type in \S\ref{s:nsmcvs}).  

\begin{figure*} \begin{center}
\includegraphics*[width=\textwidth]{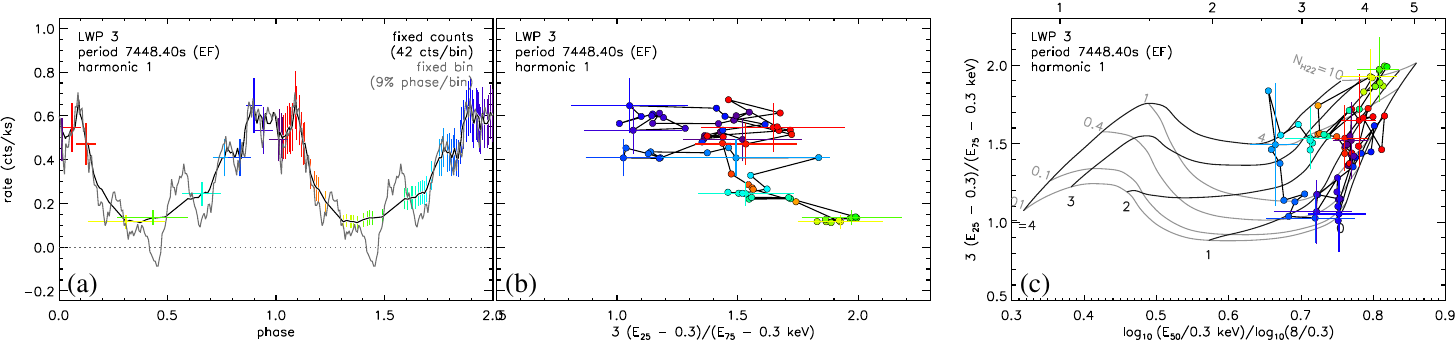}
\caption{
Phase-resolved quantile analysis of LWP~3 
using the last 801 ks exposure:
(a) the folded light curves, 
(b) the rate vs.~the quartile ratios,
and (c) the phase-resolved quantile diagram. 
The folded light curves are generated using
sliding windows of fixed-width phase bins (grey)
and variable width phase bins with fixed net counts (black: example data
points with bin sizes are shown in phase 0 -- 1)
and the error bars of the latter are color-coded
by phase for easy comparison.  
The quartile ratios is mostly
proportional to Log(\nHt) for LWP~3.}
\label{prqa}
\end{center}
\end{figure*}

\subsection{LWP~2: CXOPS J175123.5-293755} 

This source was reported as a possible accreting binary (LW~25) by B09.
Two optical sources in the \HST/ACS images are found within the error
circle of the X-ray source position, and neither of them stands out with
any unusual colors. The folded light curve shows a possibility of a
long eclipse starting at phase $\sim$ 0.5.
LWP~2 and 6 are located in the region with an excess of seemingly diffuse
soft X-ray background, which perhaps indicates the lowest extinction
region of the field. In the case of LWP~2, it shows a significantly larger
absorption in the X-ray spectrum than the field average, indicating an
intrinsic absorption in the system.

\subsection{LWP~3: CXOPS J175129.1-292924 }

This source exhibits a mild anti-correlation between the count rate and the
absorption in the X-ray spectrum according to a phase-resolved quantile
diagram (Fig.~\ref{prqa}) \citep[][H09a]{Hong04}.  
This anti-correlation is consistent with a picture that
the observed X-ray modulation is caused by the variation of intrinsic
absorption in the system, similar to the IP discovered in BW (H09a).
Alternatively if the X-ray modulation originates from an eclipse or
an obscuration of the hot spot or the emission region due to the spin
or orbital motion, then the absorption variation may not be expected to be
strongly correlated with the rate change.
The folded light curve of 
the source shows a hint of an eclipse (phase $\sim$ 0.45), which, if
true, implies the system is sychronized (i.e.~polar), considering the 
synchronized primary modulation in the folded light curve.

\subsection{LWP~4: CXOPS J175131.6-292956} \label{LWP4}

This is the brightest source of the 10 periodic sources.  The X-ray
spectrum does not show any sign of iron lines.  The BB search indicated
four independent blocks in the long term light curve, but since the
source fell near a CCD edge in Epoch 3, only two blocks can be credited
to be independent.  In summary, the observed X-ray flux in 2008 is higher
by a factor of $\sim$ 1.8 than that in 2005.

\subsection{LWP~5: CXOPS J175133.9-292754} \label{LWP5}

This source and LWP~8 were observed at relatively large
off-axis angles.

\subsection{LWP~6: CXOPS J175118.7-293811} \label{LWP6}

\begin{figure*} \begin{center}
\includegraphics*[width=0.84\textwidth]{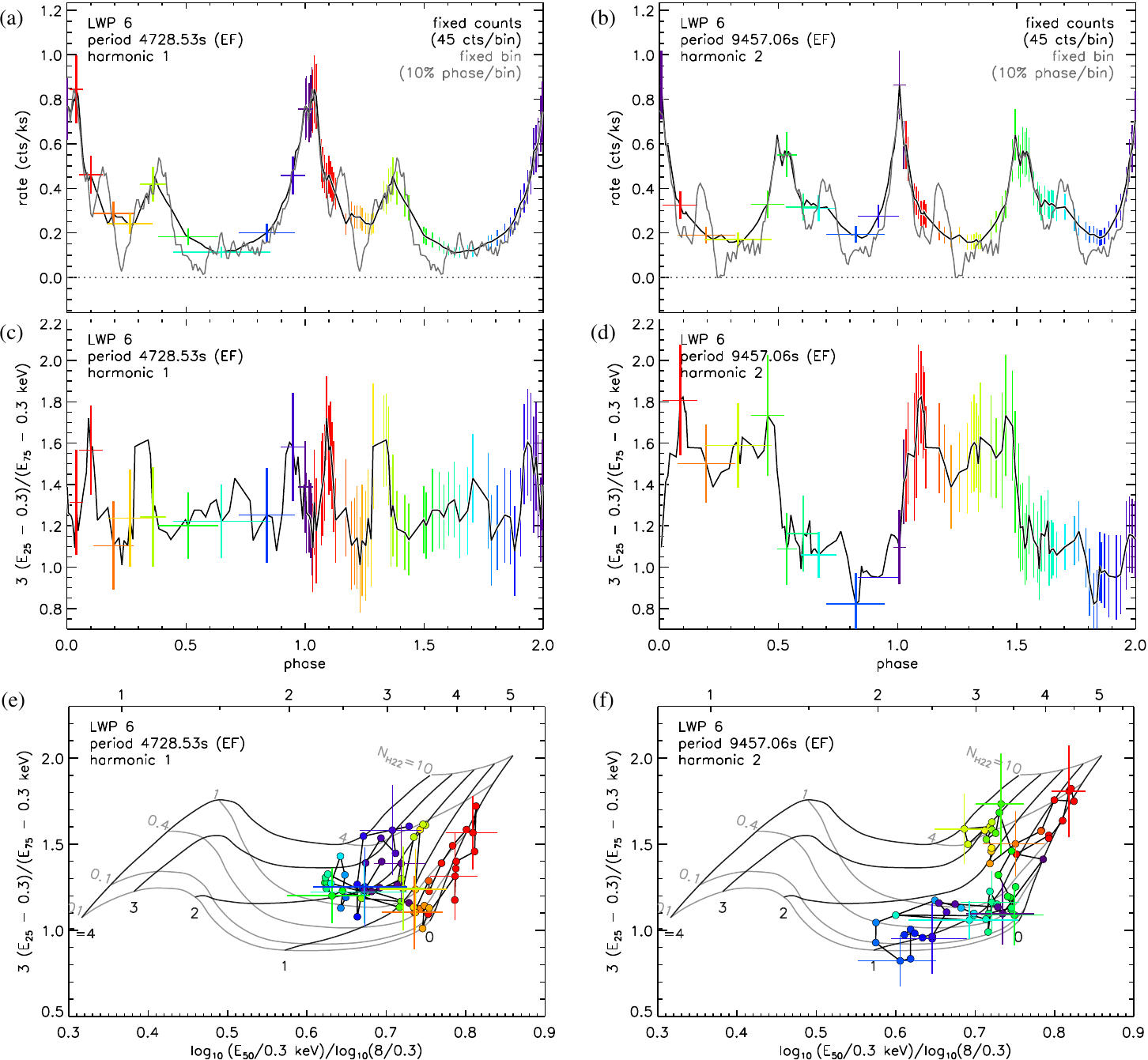}
\caption{Phase-resolved quantile analysis for LWP~6: 
the left panels for the primary period (4731 s) and the right panels for
the second harmonics (9461 s).  (a,b) the folded light curves, (c,d) the
phase-resolved estimates of \nHt using the power law models
in the quantile diagram and (e,f) the phase-resolved quantile diagram.
For the 9461 s period (d), the first half of the phase (0.0--0.5)
shows a higher absorption (\nHt $\gtrsim$ 2) than the second half
(0.5--1.0) (\nHt $\lesssim$ 1).}
\label{prqab}
\end{center}
\end{figure*}

This source was reported as a candidate CV (LW~8) by B09, based
on the high X-ray-to-optical flux ratio and H$\alpha$ excess (see Fig.~4
in B09).   The folded light curve in Fig.~\ref{basic1} shows a hint
of an eclipse at phase $\sim$ 0.2.  
Our routine analysis of phase-resolved quantiles at the primary period
and the second harmonic revealed
an intriguing result as illustrated in Fig.~\ref{prqab}.  At each
period, we show two versions of folded light curves using sliding phase
windows: one with fixed-width phase bins (10\%, grey), and the other
with variable-width phase bins but fixed net counts in each phase bin
(45 counts, black). The former is better suited to identify a sudden drop
in the count rate such as an eclipse, and the latter is better suited
to reveal a sudden increase. In addition, the latter is better suited
for phase-resolved spectral quantile analysis since every phase bin
contains enough events to allow a reliable estimate of energy quantiles.
The folded light curves in fixed-width phase bins show an eclipse-like
feature both at the primary period (phase $\sim$ 0.2 and 0.6) and at the
second harmonic ($\sim$ 0.25).  The ingress and egress of the eclipse are
sharper at the second harmonic than the primary period.  A narrow eclipse,
synchronized with the primary modulation, suggests the system is a polar.

What is interesting in this source is a spectral change correlated with
the phase.  At the primary period, the first quarter of the phase shows
an intrinsically harder X-ray spectrum ($\Gamma < 1$) than the rest
($\Gamma > 1$), which implies two different emission regions or
mechanisms are present (Fig.~\ref{prqab}e).  
On the other hand, at the second harmonic, there
appears to be a dramatic change in the absorption between the first
($\nHt\gtrsim2$) and second halves
($\nHt\lesssim1$) of the phase (Fig.~\ref{prqab}d \& f).  The 
change in the absorption between two peaks of the folded light curve
would suggest that two magnetic poles of the system are visible in
turn, and the X-ray emission from one of the two undergoes 
through more material (likely an accretion curtain, trailing to a pole and
extended from about half of the accretion disk or ring) before
reaching us.  Fig.~5 in \citet{Evans07} illustrates a
possible viewing geometry for such a system: unlike their exmaples,
the soft blackbody component ($\ll$ 1 keV)
of LWP~6 is likely always invisible due to the interstellar absorption,
but the geometry allowing for a phase-dependent variation of intrinsic
absorption due to an accretion stream or curtain can apply to LWP~6.
This picture, if true, suggests that the second harmonic (9457 s) is a
real period and the observed primary period (4729 s) is a sub-harmonic, even though
the LS method did not find the second harmonic significant.
Further observation is required to determine which period represents
the true orbital and spin geometry. In addition, a new diagnosis may be
required to quantitatively evaluate the significance of various types of
spectral changes \citep[see][]{Connors11}.

\subsection{LWP~7: CXOPS J175122.7-293436} \label{s:lwp7}

During three pointings out of 8 total in Epoch 4, this source fell near
a CCD edge (see the effective area in Fig.~\ref{basic2}),
which would discredit the observed periodicity, but the relatively
clean data set (the last 352 ks GTI) free of near-CCD-edge events also
exhibits a significant periodicity at 12076 s. The net count
of LWP~7 in the last 352 ks GTI is only 119, which is similar to that
of J175055.5-292948 ('+' in Fig.~\ref{f:rundown}) in Epoch 4, but $P$\Ss{det} of
LWP~7 is estimated substantially higher than J175055.5-292948 because of the
relatively lower background counts in the source aperture region of LWP~7.
Therefore, we consider the observed periodicity of LWP~7 valid.

\subsection{LWP~8: CXOPS J175147.4-294215}

The folded light curve of the source shows an eclipse-like feature at
phase $\sim$ 0.6.

\subsection{LWP~9: CXOPS J175133.6-293313} \label{s:lwp9}

LWP~9 was reported as a potential accreting binary (LW~19) by B09.
The aperture source region (95\% PSF) of these sources mildly overlaps
with that of a neighbor.  A clean data set relatively free of 
contamination from the neighbor (see H05 for aperture choice) exhibits
the same periodic modulation, so that its periodicity is considered valid.

\begin{figure*} \begin{center}
\includegraphics*[width=0.98\textwidth]{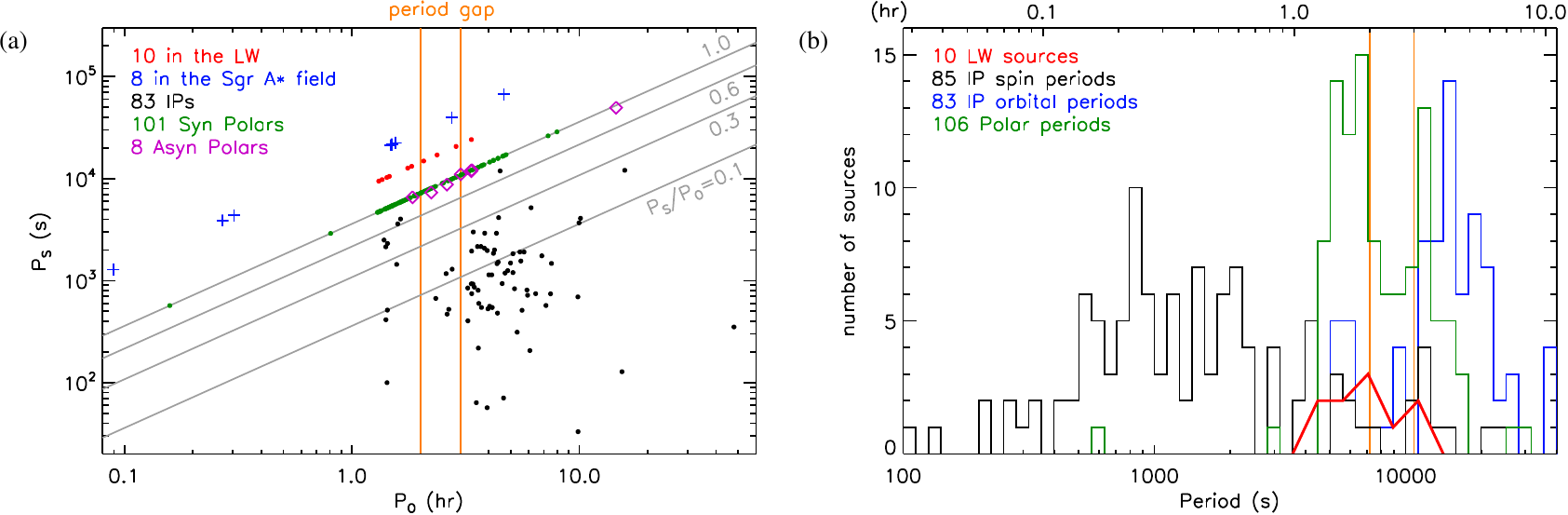}
\caption{(a) The spin (\Ps) vs.~orbital period (\Po) distribution of
MCVs from the RK catalog (ver~7.15). The vertical yellow lines mark
the period gap, and the diagonal lines represent \Ps/\Po = 1,
0.6, 0.3 and 0.1.
The purple diamonds show eight asynchronous polars (APs) (three are on
top of each other), and there are six 
near synchronous IPs with \Ps/\Po $>$ 0.3.
The observed periods of the periodic sources in
the LW (red) and the Sgr A* field (blue) are marked as orbital periods,
along \Ps = 2 \Po or 4 \Po for clarity (those less than an hour are
likely the spin periods).
(b) The period distribution of the LW sources (red) is compared to
the distribution of the spin (black) and orbital (blue) periods of IPs
and the periods (green) of polars in the RK catalog. The periodic
sources (red) in the LW fill in the period gap, indicating these
are likely MCVs.  The periods of the LW sources are distributed more
closely to those of polars (green) than the spin (black) or orbital
(blue) periods of IPs. }
\label{pdist}
\end{center}
\end{figure*}

\subsection{LWP~10: CXOPS J175119.4-293659} \label{s:lwp10}

According to the EF analysis for LWP~7, 9
and 10, a higher harmonic exhibits more significant periodicity than
the primary period found by the LS method. While the differences in
the significance between the primary periods and their harmonics are
marginal for two sources, in the case of LWP~10 the second harmonic
appears substantially more significant. The $\chi^2$ of the 15-bin
folded light curve with respect to the constant rate is 133 at the second
harmonic vs.~89.5 at
the primary period\footnote{For comparison, the expected $\chi^2$ of
the 15-bin folded light curve is 70.9 at \fap = 1\% for $N_S$= 381 and
$N_P$=19911, although the EF search was conducted only for the 10 sources
at 4000 periods around the likely true modulation periods.}.  Therefore we
consider the second harmonic to be the real modulation period for LWP~10.
The high $\chi^2$ value at the second harmonic is partially due to an
eclipse-like feature in the folded light curve at phase $\sim$ 0.7,
which suggests the system is a polar.

\section{Discussion} \label{s:dis}

In this section, we investigate the most probable source types for
the periodic sources found in the LW and their implications, based on
the statistical distribution of their properties.  We also explore the
hidden population of periodic sources in the Bulge X-ray sources in the
GCR through completeness simulations for periodicity detection.

\subsection{Unusual MCVs?} \label{s:nsmcvs}

The observed source
properties such as the X-ray luminosity range
($\sim$ 10\sS{30-33} erg s\sS{-1} for distance of 1 -- 8 kpc), the
relatively hard X-ray spectra ($\Gamma <$ 2), the period range
(1.3 -- 3.4 hr), and the relatively high X-ray-to-optical flux ratios,
all indicate that these sources are typical MCVs.  But the collection of
these properties does not appear to fit well with most common types
of MCVs as explained below.  

MCVs can be largely divided into two groups, IPs and polars, depending
on the relative strength of the magnetic fields.  Traditionally, polars
are synchronized or nearly synchronous (\Ps/\Po $\sim$ 0.98 -- 1.02),
whereas IPs are not (\Ps/\Po $\lesssim$ 0.1).
Fig.~\ref{pdist} shows the spin vs.~orbital period distribution
of the MCVs in the Ritter \& Kolb (RK) catalog (ver~7.15) \citep{Ritter03}.
In Fig.~\ref{pdist}a, the observed periods of the periodic X-ray sources
found in the LW and Sgr A* fields are shown as orbital periods 
along \Ps/\Po = 2 or 4 for easy comparison.

In the case of the periodic sources in the LW, the observed period
distribution (red in Fig.~\ref{pdist}b) resembles those of polars
(green) better than either the spin (black) or orbital (blue) periods
of IPs.  Some IPs do have spin or orbital periods at around an hour
to three hours, but the majority of the spin (or orbital) periods are
shorter (or longer), whereas the majority of the periods of polars are
in the same range as those of the periodic sources in the LW.

In order to find out if the above result is due to a period-dependent
selection bias in the periodicity search routines, we have conducted a
set of simulations using synthetic light curves with various net counts
(100 to 1000), modulation amplitudes ($A_0$ = 10\% to 100\%) 
and periods ($\sim$ 20 to 1.1\x $10^{5}$ s).  For each combination
of parameters, we generate 500 synthetic light curves, which allows
$\lesssim$ 2\% accuracy in measurement of the detection probability.
As in the simulations in \S\ref{s:srcsel} (and \S\ref{s:complete}),
each synthetic light curve is generated to properly reflect the GTIs
and Barycentric shifts of the CCD readout time of the real data.

\begin{figure} \begin{center}
\includegraphics*[width=0.45\textwidth]{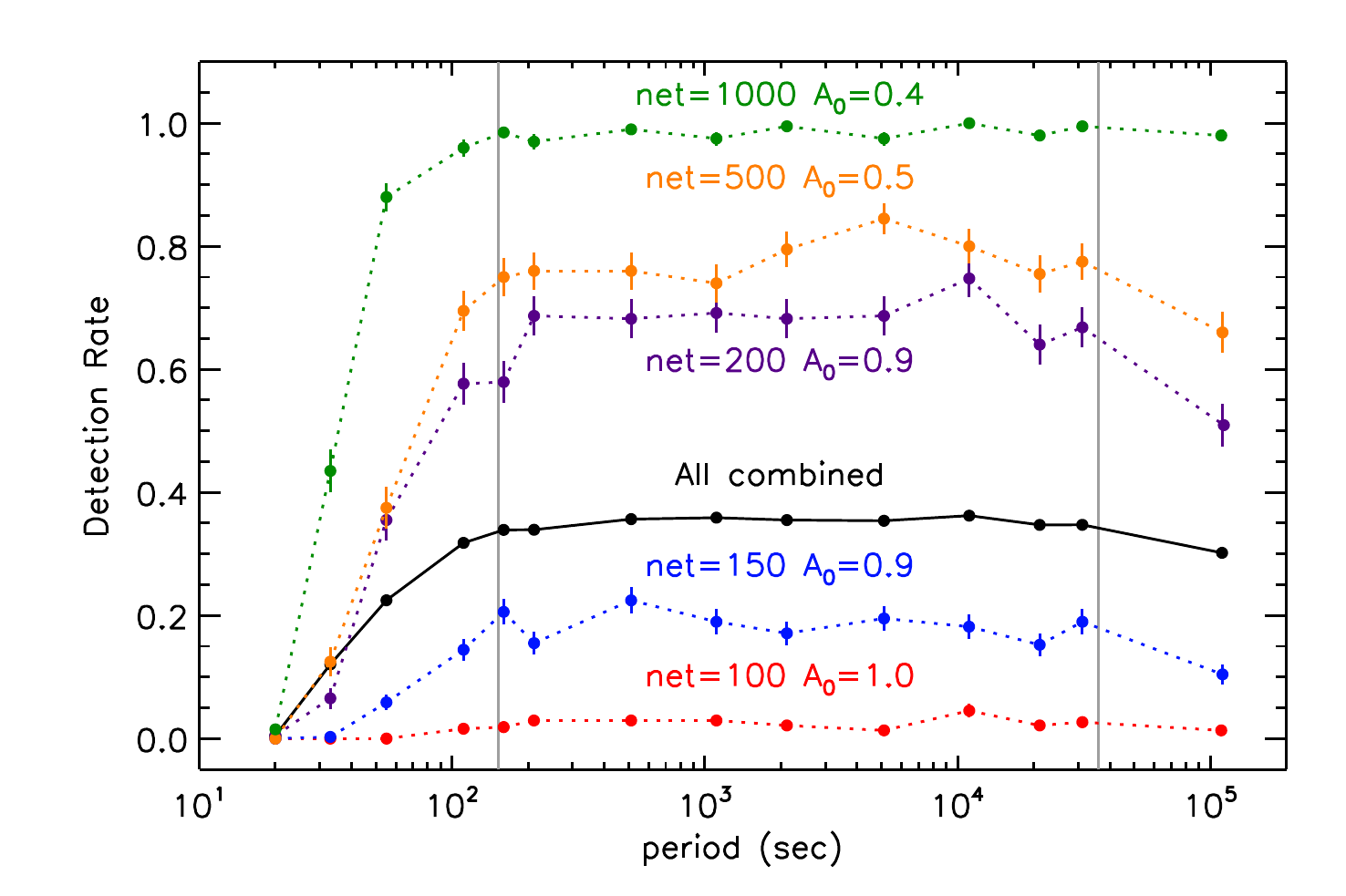}
\caption{Examples of detection (\fap (Fix) $<$ 1\%)
probabilities of periodicity as a function of period by
simulations. The plot for "All Combined" includes the following cases:
for net = 1000, $A_0$ = 0.2 .. 0.5, for net = 500, $A_0$ = 0.3 .. 0.7,
for net = 200, $A_0$ = 0.5 .. 1.0, for net = 150, $A_0$ = 0.6 .. 1.0,
and for net=100, $A_0$ = 1.0. The detection probability show no bias
within the period range
of the interest between 150 s to 10 hr (two vertical grey lines).
}
\label{f:pbias}
\end{center}
\end{figure}

The simulation results show there is no significant
selection bias in the range of 150 sec to 10 hr (Fig.~\ref{f:pbias}).
This implies that the resemblance to the polar period distribution is
not due to any selection effect in
the search algorithms\footnote{There is, however, a selection bias towards
sources with high modulation amplitudes, as expected. See
\S\ref{s:complete}.}, or if these sources are IPs, they indeed belong
to a statistically different population from the typical IPs in the
RK catalog.  In addition to the observed eclipses (or eclipse-like
features) synchronized with the primary modulation of the pulsed profiles
(e.g.~LWP~3, 6, 8 or 10), missing secondary periods from all the
sources in Table~\ref{t:detection} indirectly supports
the systems being polars - synchronized systems, although
non-detection does not guarantee the absence of the secondary periods.

\begin{figure} \begin{center}
\includegraphics*[width=0.49\textwidth]{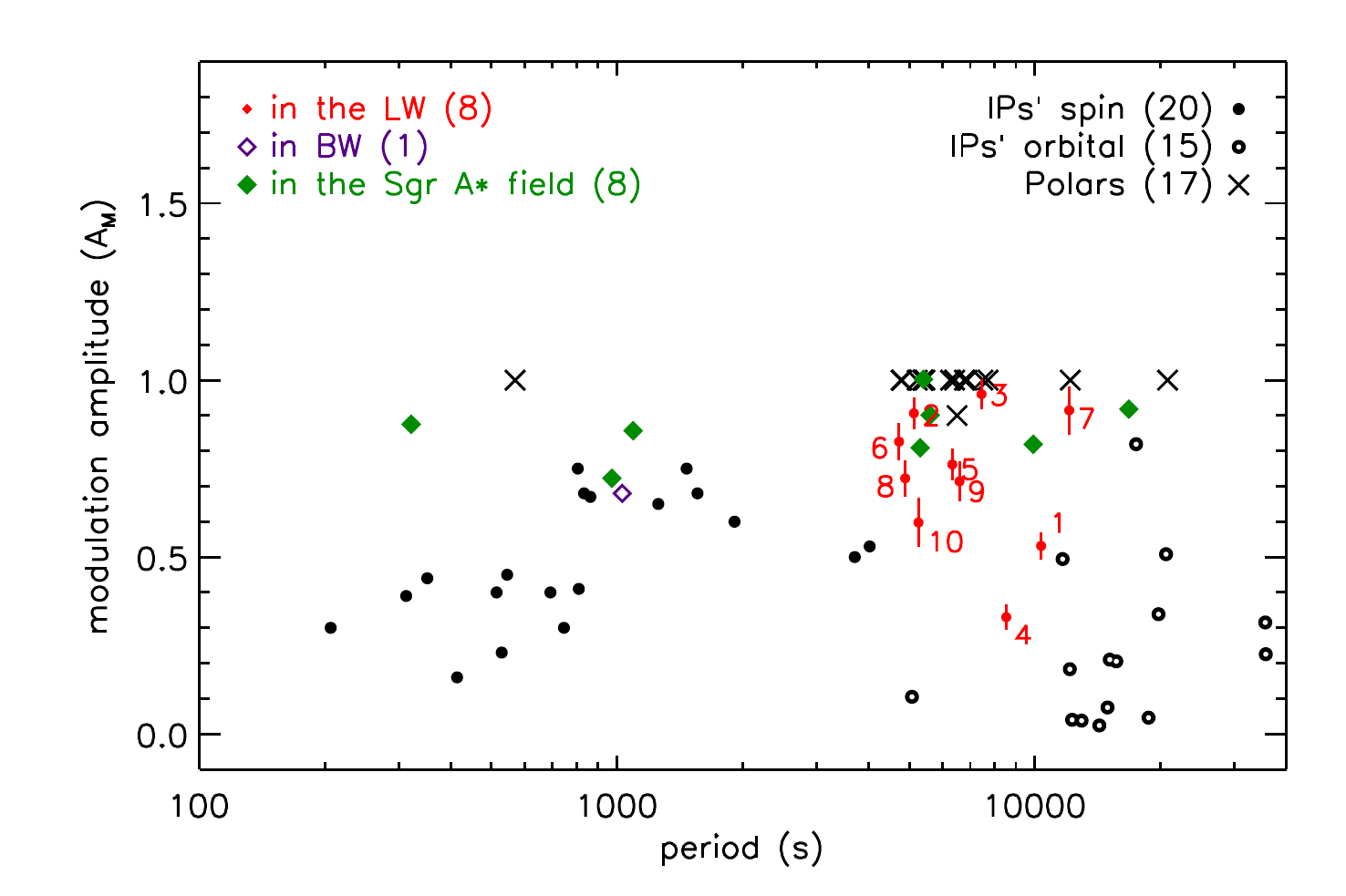}
\caption{
The X-ray modulation ($A_M$) vs.~the period distribution 
of literature-selected IPs (black closed circles for spin
periods and black open circles for orbital periods),
polars (black 'x's), 
and the periodic sources
in the LW (red circles), BW (purple diamond) and the Sgr A* field 
(green diamonds). 
[References] 
Orbital periods of IPs: \citet{Parker05}. 
Spin periods of IPs:
GK Per, HT Cam, EX Hya, RX J1548.2, 
AO Psc, V1223 Sgr, AE Aqr -- \citet{Evans05}, 
V405 Aur -- \citet{Evans04a},
FO Aqu -- \citet{Evans04b},
PQ Gem -- \citet{James02},
IGR J15094 -- \citet{Butters09},
UU Col -- \citet{Martino06},
WX Pyx -- \citet{Schlegel05},
XY ARI -- \citet{Salinas04},
T Leo -- \citet{Vrielmann04}, 
TV Col -- \citet{Rana04},
V1062 Tau -- \citet{Hellier02a},
1WGA J1958.2 -- \citet{Norton02},
YY Dra -- \citet{Szkody02},
V709 Cas -- \citet{Norton99}. 
Polars:
V1309 Ori -- \citet{Schwarz05},
EK UMa -- \citet{Beuermann09},
HU Aqr -- \citet{Schwarz09},
V2301 Oph -- \citet{Ramsay07},
SDSS J015543 -- \citet{Schmidt05},
EP Dra -- \citet{Ramsay04c},
OY Car -- \citet{Wheatley03},
DP Leo -- \citet{Ramsay01},
RX J1846.9 -- \citet{Schwarz02},
V407 Vul -- \citet{Marsh02},
CE Gru -- \citet{Ramsay02},
V1432 -- \citet{Rana05},
V347 Pav, GG Leo, EU UMa -- \citet{Ramsay04a},
RX J1002-19 -- \citet{Ramsay03})
}
\label{pedmod}
\end{center}
\end{figure}

Fig~\ref{pedmod} illustrates a wide range of modulation
amplitudes of the 10 periodic sources in the LW.
For comparison, we also show some of literature-selected IPs and polars.
If the X-ray emission originates from a small spot (e.g. polar cap)
on the WD surface, the modulation due to the 
spinning motion of the
compact object is expected to exhibit a larger amplitude change than
that from the orbital motion. 
For instance, a sample of IPs in
Fig.~\ref{pedmod}, selected from the literature, show a slightly
higher average value of the modulation amplitude at the spin
periods (black closed circles) than at the orbital periods (black
open circles). However, the modulation distributions of 
these IPs and polars in
Fig.~\ref{pedmod} are likely selection biased: e.g.~17 polars in
Fig.~\ref{pedmod} are mostly eclipsing systems, which are likely
preferentially found in periodicity searches due to the large modulation
amplitude. Note for polars in Fig.~\ref{pedmod},
$A_M$ is simply based on $R\Ss{min}$ and $R\Ss{max}$ of the published
light curves, where by definition, $A_M$=1 for eclipsing systems, whereas
for the periodic sources in the LW, we calculate $A_M$ from $A\Ss{rms}$
(e.g.  for LWP~1, $A_M$=0.53 instead of 1.0).  In summary, it is not
easy to link the distribution of the modulation amplitudes to a particular type
of MCVs due to the lack of systematic survey. However, even in
Fig.~\ref{pedmod} with selected samples of IPs and polars, the period distribution of the
10 periodic sources strongly favors polars.

\begin{figure*}[t] \begin{center}
\includegraphics*[width=0.98\textwidth]{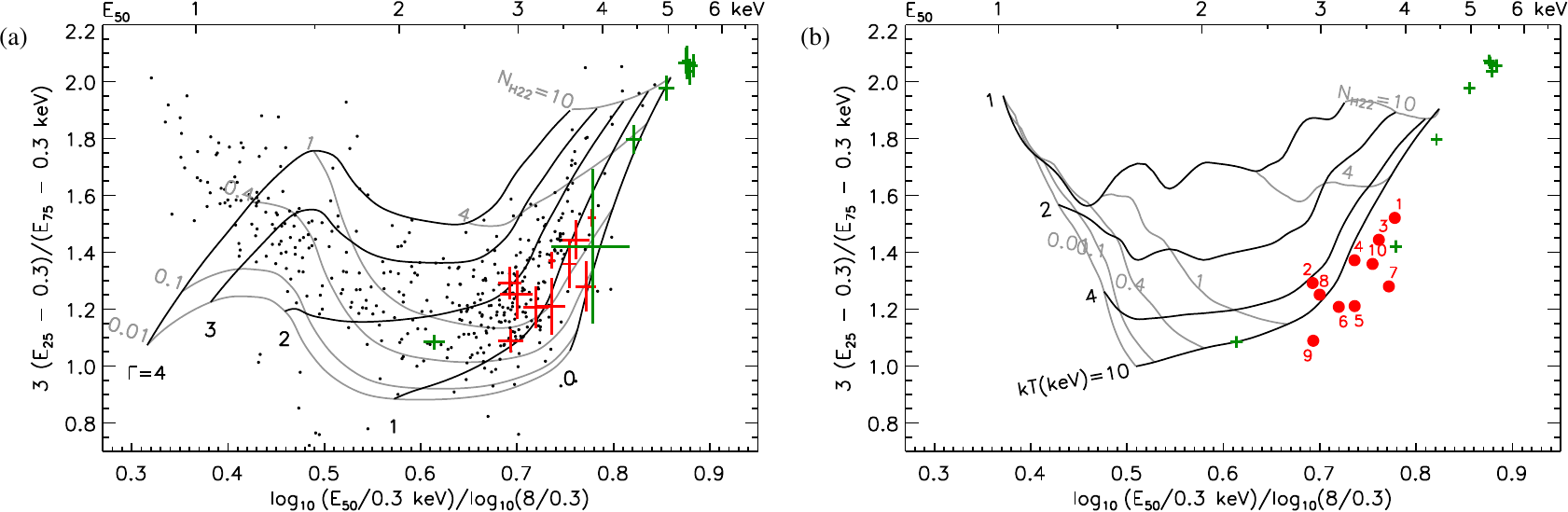}
\caption{Quantile diagrams (H04, H09a) of the periodic X-ray 
sources in the LW
(red) and the Sgr A* field (green). 
The same data points are plotted
over a set of power model grids (a) or APEC
model grids (b). 
The (black) dots in (a) are the discrete
X-ray sources in the LW with net counts greater than 50 in the 0.3--8
keV band.
The model grids are based on the response function at
the aimpoint, which can be partially responsible 
for a small discrepancy 
between the spectral fit and quantile analysis
in the parameter estimates for some sources
(e.g.~in LWP~9, \nHt = 0.2(3) from the spectral fit
vs.~0.7(2) from the quantile analysis).  
}
\label{qd}
\end{center}
\end{figure*}

Unlike the period distribution, the relatively hard X-ray spectra of
the periodic sources in the LW imply that these systems are likely IPs.
The quantile diagrams in Fig.~\ref{qd} illustrate
the overall spectral hardness of the periodic sources (red) in comparison with
the rest of the X-ray sources (black dots, net counts $\ge$ 50) in the LW.
Fig.~\ref{qd}a and b
overlay a set of simple power law and APEC model grids
respectively over the same data points.  
The eight periodic
sources found in the Sgr A* field
are also shown in (green) crosses (M03b).  All of the periodic
sources show an intrinsically hard X-ray spectrum, similarly to the periodic
sources in the Sgr A* field.

IPs tend to show harder X-ray spectra, which are
associated with higher accretion rates and weaker magnetic fields
\citep{Ramsay04d}.
For instance, in the case of polars, the X-ray spectra are well described
by a blackbody component with $kT<60$ eV and a two-temperature thermal
plasma component with $kT_1=0.7-0.9$ keV and $kT_2=3-5$ keV
\citep[e.g.][]{Ramsay04b}, whereas the X-ray spectra of IPs show a blackbody
component with $kT>60$ eV and a one or two-temperature thermal plasma
component with $kT\ge 10$ keV \citep[e.g.][]{Anzolin08}.  Due to the
interstellar
absorption in the LW field (\nHt$\sim$ 0.7), the blackbody
component is usually undetectable, but the spectral distinction of the plasma
components between polars and IPs remains detectable in the \chandra X-ray
band: in a quantile diagram, polars would lie in the upper-left section of
$kT\sim 4-10$ keV line, and IPs in the lower-right section
(Fig.~\ref{qd}). Although the above description of the X-ray spectra of
IPs and polars is over-simplified and without a systematic survey (e.g.~LWP~6
is likely a polar, see \S\ref{LWP6}), it is
generally accepted that IPs exhibit a harder spectrum.  For instance,
a recent survey of hard X-ray sources ($\gtrsim $ 15 keV) conducted
by \swift/BAT and \integral/IBIS shows that the composition of MCVs in the
hard X-ray band are predominantly IPs \citep{Scaringi10}.

The period range and the relative hard X-ray spectra of the LW periodic sources
also resemble those of some of the periodic sources found in the Sgr
A* field (green data points in Fig.~\ref{qd} and blue points in
Fig.~\ref{pdist}a) (M03b).  Of eight periodic sources in the
Sgr A* field, four are in
the same period range as the LW periodic sources with a large modulation
amplitude ($A\Ss{RMS}>$40\% or $A_M>$72\%), and all of them show
relatively hard X-ray spectra. One of
them was identified as a foreground polar, based on the light curve, and
it lies at the line of $kT = 10$ keV and the rest lie in the $kT > 10$
keV section.  
In the case of the Bulge X-ray sources in the Sgr A*
field, the heavy interstellar absorption ($\sim$ 6\x 10\sS{22} cm\sS{-2}) 
may complicate identification of the intrinsic X-ray spectra, but for
the periodic sources in the LW ($\sim$ 7\x 10\sS{21} cm\sS{-2}), there
is little doubt that most of them exhibit intrinsically hard X-ray spectra.

An obvious question, then, is, what is the nature of the periodic X-ray
sources found in the LW and the GCR, whose period distribution resembles
polars' but X-ray spectra resemble IPs'?  

First, one can consider these are a rare type of polars with unusually hard
X-ray spectra. Although uncertain, it is speculated that the origin of harder
X-ray spectra in IPs relative to polars is related to the weaker magnetic field
strength and the subsequently deeper penetration of the accretion stream
into the WD surface.  For instance, \citet{Cumming02} suggested the relatively high
accretion rate in IPs effectively buries the WD magnetic field, making
them appear less magnetic, which in turns helps maintaining the high
accretion rate \citep[e.g.~V407 Vul: see][]{Marsh02}.  This idea nicely
ties the dichotomy of the magnetic field strength between IPs and
polars. Under this picture, one can imagine some unusual evolutionary
scenarios from IPs to polars, where polars survive with a high accretion
rate that can bury the magnetic fields, allowing a harder X-ray spectrum.
If most of these periodic sources are located near the Galactic center
(8 kpc) as expected from the high absorption in their X-ray spectra
and the high stellar density of the Bulge, the
X-ray luminosities of these sources
are estimated at the high end
of the MCV range ($\gtrsim 10^{32}$ erg s\sS{-1}, see
Table~\ref{t:spec}), which is also consistent
with the above picture.

Second, a rare subclass of
MCVs, nearly synchronous MCVs (ns-MCVs), perhaps also meet both of the observed
properties - the period distribution and the relative hard X-ray spectra of
the periodic sources in the LW.  One can divide ns-MCVs in two subgroups
- nearly-synchronous IPs (ns-IPs) and asynchronous polars (APs). Both
subgroups may exhibit similar X-ray properties, but probably represent
different stages in the evolutionary path of MCVs.

First, APs, consisting originally of just four systems, which recently
extended to eight according to the latest RK catalog (ver 7.15), are traditionally
considered as polars that are temporarily out of synchronization due to
a recent Nova activity, which has altered their
magnetic locking, giving \Ps/\Po$\sim$0.98--1.02.  Interestingly it
is speculated that APs exhibit a harder spectrum than normal polars,
but with the similar periods, as marked with purple diamonds in
Fig.~\ref{pdist}a. For instance, two
of seven APs as opposed to two of 92 normal polars are found in the hard
X-ray survey using \integral/IBIS, \swift/BAT and \suzaku/HXD
\citep{Scaringi10}.

Second, there are increasingly more IPs found near at synchronization
(\Ps/\Po$>0.3$).  Starting with EX Hya, the list increases to six
according to the RK catalog. Their orbital periods are predominantly
around 1.5 hr except for V697 Sco with a 4.5 hr orbital period.
According to the evolutionary model of \citet[hereafter
N08]{Norton08}, IPs start out with \Ps/\Po $<0.1$
and as the systems evolve through magnetic lock, the orbital periods
decrease and the spin periods increase, i.e. \Ps/\Po approaches 1.
Therefore, the orbital periods of ns-IPs will be clustered around or
below the period gap near the end of the evolution, resembling
the period distribution of polars more closely than that of usual, unsynchronized IPs.

As with polars exhibiting unusally hard X-ray spectra, the presence of ns-MCVs
is very intriguing in terms of the evolutionary models of MCVs, challenging the
conventional view of IPs with \Ps/\Po $\lesssim$ 0.1 and polars $\sim$
0.98 -- 1.02.  For instance,  Paloma or RX J0524+42
\citep{Pineault87}, recently identified
as an AP, shows \Ps/\Po $\sim$ 0.93, and its relatively 
large de-synchronization
($\sim$ 7\%) compared to the conventional APs ($<$ 2\%) suggests that
this system might represent
the missing link of the evolutionary path between IPs and polars
\citep{Schwarz07}, rather than polars being out of synchronization
temporarily due to recent Novae activities.  As shown in
Fig.~\ref{pdist}a, the gap ($\sim$ 0.3 -- 0.95) in the period ratio
between IPs and polars are now more or less bridged by the recent 
discoveries of many ns-MCVs. It is speculated that some of these ns-MCVs
transit between IPs and polars \citep[e.g.~V1025 Cen, see][]{Hellier02b}.

According to N08, if the period ratio exceeds 0.6, the only stable
equilibrium is at synchronization (\Ps/\Po =1).
\citet{Schwarz07} suggest that the probability of finding ns-MCVs is
very low, considering relatively short timescale for synchronization ($<$
1 Myr) compared to the lifetime of a CV ($\sim$ 100 Myr) (see also N08).
Therefore, if many of the periodic sources in the LW are indeed ns-MCVs,
it imposes another constraint on the evolutionary model or suggests an
unusual environment of the Galactic Bulge capable of harboring many such
rare systems.   The similar statement can be made for polars with
unusally hard X-ray spectra.

The relative composition of source types in MCVs are
highly biased, depending on search wavelengths. In the RK catalog, where
most of the discoveries are based on optical/UV or longer wavelength
bands, the relative ratio of IPs vs.~polars
are close to 1 (e.g. 83 IPs vs.~109 polars in Fig.~\ref{pdist}),
whereas a hard X-ray survey ($\gtrsim$ 15 keV) in
\citep{Scaringi10} revealed 37 IPs and only 2 polars (both are APs).
Therefore, if many of the periodic sources in the LW are indeed unusual polars
or ns-MCVs,
it indicates the \chandra X-ray band is well tuned for 
discovery of these rare MCVs.

\begin{figure*} \begin{center}
\includegraphics*[width=0.99\textwidth]{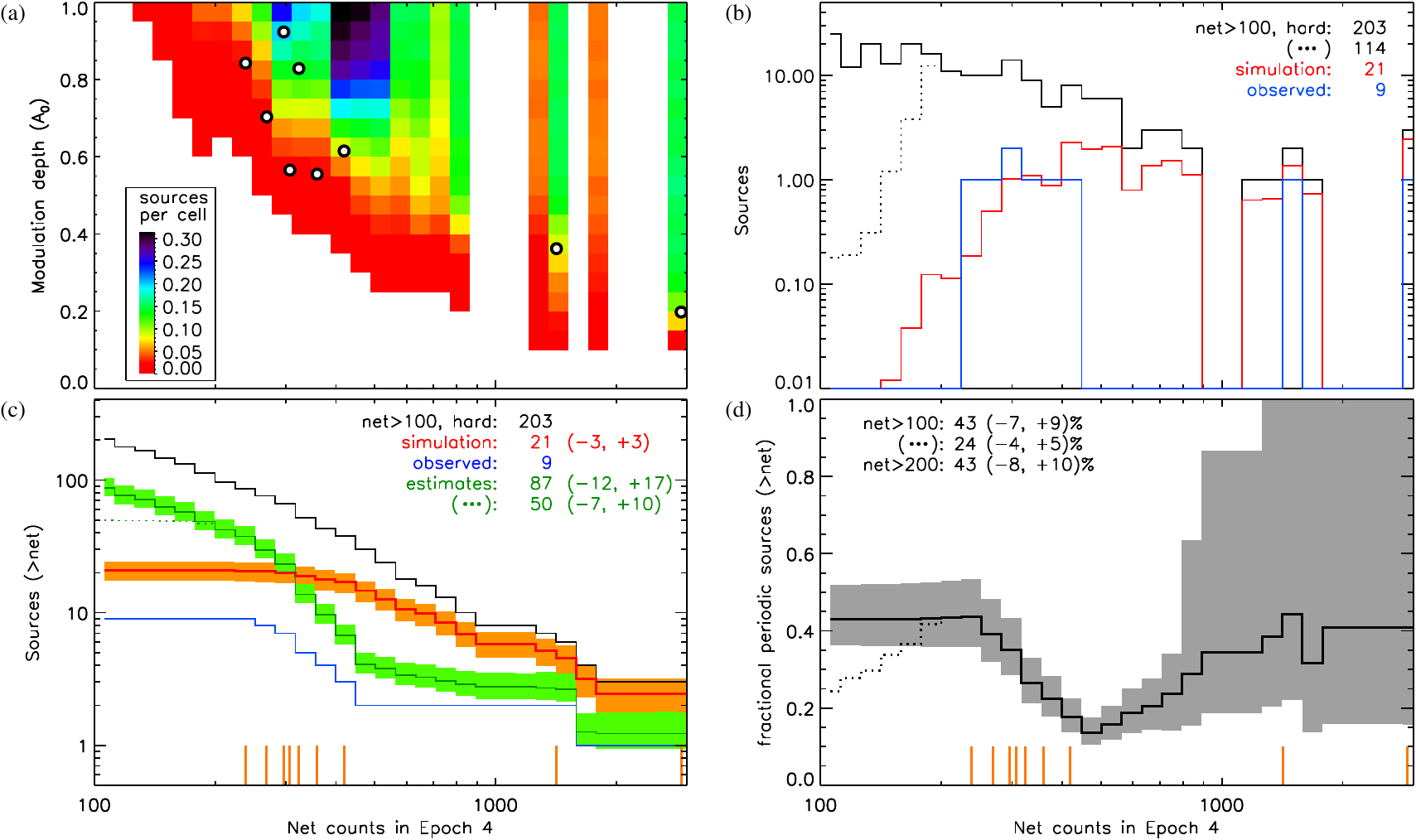}
\caption{Completeness study for periodicity detection 
for hard X-ray sources ($E\Ss{50} \ge 2.5$ keV; 
see also case 3 in Table~\ref{simdet}).
(a) The (relative) expected distribution of detectable periodic sources based on
the completeness simulation under the assumption of all the sources being periodic with
the uniform distribution of modulation amplitudes.
The open circles show the 10 periodic sources in the LW.
(b) The number of the sources with detectable periodicity (red) as a function
of net count if all the sources (black) are periodic in
comparison with the observed periodic sources (blue). The dotted line
indicates the testable sample of the population if we limit
the detection sensitivity at 1\%.
(c) The accumulated number of the detectable sources (red) if all the
sources are periodic (black). Given the observed periodic sources (blue),
we also show the expected total population of periodic sources (green). Below
200 counts, the detection sensitivity is too low to
make a reliable assessment of the hidden population of periodic sources
(dotted green lines). 
The shade for green and blue indicates the statistical errors of the
simulation results and estimates.
(d) The fractional percentage (solid line) of the periodic sources, which is the ratio
of the green to black lines in (c). The dotted line shows a conservative limit
below 200 counts due to lack of detection sensitivity, but the rapid rise
of the percentage of periodic sources as the count approaches from 500 to 200 indicates
the trend will continue at low counts, where the majority of the Bulge
MCV populations lie (see \S\ref{s:complete}).
}
\label{detprob}
\end{center}
\end{figure*}

Finally we note that some of the observed X-ray properties of the periodic X-ray
sources in the LW are shared by a group of quiescent Low Mass X-ray
Binaries (qLMXBs).  For instance, Swift J1353.5-0127, recently observed
in an outburst as a black hole (BH) transient,
may have a relatively short orbital period \citep[$\sim$ 2 hr, see][]{Casares11}
and exhibit a hard X-ray spectrum \cite[][]{Krimm11}.  
The observed over-abundance of X-ray transients within 1 pc of Sgr A*
\citep[cf.~4 out of 7 within 20 pc, see][]{Muno05} implies  that a large
number of qLMXBs may be present as dark stellar remnants within 1 pc of Sgr A*
\citep{Schodel07}.  However, CVs (and MCVs) are expected dominantly more abundant
than these BH transients, and the periodic X-ray sources in the LW do
not exhibit any strong outbursts in the 1 Ms exposure spanned over 3
years \citep[cf. out of 7 X-ray transients in the Sgr A* field, 3 or 4
sources have been observed in outbursts in each year,
see][]{Degenaar09,Degenaar10}
it is reasonable to think that
the large fraction of the periodic sources in the LW are in fact MCVs.

\begin{table*}
\small
\caption{Estimation of the total number of periodic sources in the LW
by completeness simulations}
\begin{tabular*}{\textwidth}{c@{\extracolsep{\fill}}lccccccc}
\hline\hline
(1)                    & \multicolumn{1}{c}{(2)}       & (3)     & (4)                                        & \mbox{(5)}    & \mbox{(6)}      & (7)             & (8)                                       & (9)        \\
Cases                  &  Source Selection             & Total   & \multicolumn{4}{c}{Periodic X-ray Sources}                                            & \multicolumn{2}{c}{Alternative Estimates}           \\
\cline{4-7}\cline{8-9} & (net $\ge$ 200 in Epoch 4)    & Sources & Observed                                   & Detectable    & Estimated Total & Percentage      & Estimated Total                           & Percentage \\
\hline
1                      & all                           & 153(12) & 9                                          & 28 ($-4$, +4) & 48 ($-6$, +8)   & 32 ($-5$, +6)   & 48(15)                                    & 32(10)     \\
2                      & non-edge                      & 113(11) & 8                                          & 21 ($-4$, +4) & 43 ($-6$, +9)   & 38 ($-7$, +9)   & 43(14)                                    & 38(12)     \\
3                      & $E_{50}\ge 2.5$ keV           & 96(10)  & 9                                          & 21 ($-3$, +3) & 42 ($-6$, +8)   & 43 ($-8$, +10)  & 42(12)                                    & 43(13)     \\
4                      & $E_{50}\ge 2.5$ keV, non-edge & 69(8)   & 8                                          & 16 ($-3$, +3) & 35 ($-6$, +8)   & 51 ($-11$, +14) & 35(11)                                    & 51(16)     \\
\hline
\end{tabular*}

(1) Case number
(2) Source selection criteria. We limit the sample to sources with $\ge$ 100 net counts
in Epoch 4.
(3) The total number of sources that meet the source selection
criteria. 
(4) The number of the observed periodic sources
from Table~\ref{t:detection}. 
(5) The estimated number of the detectable periodic sources 
by simulations under the assumption that
all the sources are periodic with a uniform
distribution of modulation amplitude. 
(6) The estimated total number of the periodic sources
in source selection: (3) \x (4) / (5).
(7) The percentage of periodic sources in the selection: (6)/(3).
(8) An alternative (error) estimate of the total number of periodic sources
and (9) their percentage using negative binomial distribution,
where, for the observed periodic sources, $N$, and
the non-detection probability, $p$, given by the simulations
(in case 3, $N=9$, $1-p=21/100$), 
the estimated unidentified periodic sources and its variance are $Np/(1-p)$
and $Np/(1-p)\sS{2}$ respectively.
\label{simdet}
\end{table*}

\subsection{Periodic source content \\ in the GCR X-ray source population}
\label{s:complete}

In this section, we estimate the total number of periodic X-ray sources in
the LW through completeness simulations for periodicity detection.
We have generated 500 synthetic light curves for a given set of X-ray
modulation parameters including modulation period and amplitude, 
net count and background count.  Then we run the same detection algorithm
to see how often the synthetic light curves are detected as periodic.
This calculates $P\Ss{det}$ for the given set of parameters.
Here "detection" is simply defined as \fap (Fix) $<$1\%\footnote{The result
in this section is consistent with that acquired with "detection"
defined as \fap (Var) $<$1\%.}.
The simulation is designed to match the data set observed in Epoch 4
with the same GTI gaps, and the simulated time tags
are generated according to the Barycenter
corrected CCD readout time cycles.
The initial set of simulations indicates that the detection results
do not depend on the given period range (100 s to 10 hr)
(Fig.~\ref{f:pbias} in \S\ref{s:nsmcvs}), so we
fix the period at 5432.1 sec and varied the rest of the parameters.
The simulated modulation amplitude ($A_0$) ranges from 10 to 100\% in 10\%
increments. The simulated net counts of sources are 50, 100, 150, 200, 500,
and 1000, 3000, and the background counts are 30, 50, 100, 200, 500,
and 1000. 

Using these simulation results, which cover most of the parameter space
for the sources found in the LW, we estimate the periodicity detection
probability ($P\Ss{det}$) of the 381 sources with net counts $\ge$ 100 in
Epoch 4.  Since we do not know a priori the distribution of modulation
amplitude and the observed periodic sources show a wide range of the
modulation amplitude ($A_0$), we randomly assign a modulation amplitude
($A_0$) to each source, assuming a uniform distribution from 0.1 to 1.
We repeat the interpolations 10000 times (i.e. 10000 set of simulations),
each with random assignments of modulation amplitude to cover the full
modulation range for every source.

Fig.~\ref{detprob} shows an example result,
using 203 sources with net counts $\ge 100$ and
$E\Ss{50} \ge 2.5$ keV (hard X-ray sources, see also case 3 in Table~\ref{simdet})
under the assumption that these sources are all periodic with a uniform
distribution of the modulation amplitude.
Fig.~\ref{detprob}a shows the expected number of sources with detectable
periodicity in the 2-D phase space of net counts vs.~modulation depth
for hard X-ray sources.  The absolute number in
each cell of the phase space depends on the cell size, so what matters
here is a relative variation from cell to cell and the total number.
The result shows that we expect to detect periodicity of 21(3)
sources if all of the 203 sources are periodic with a uniform distribution
of modulation amplitude.

Fig.~\ref{detprob}b \& c show the same result (differential and accumulated)
as a function of net count.
The source distribution with detectable periodicity (red) indicates
that the detection probability for sources with net counts less than 150 
is very low\footnote{This is an over-simplification since $P\Ss{det}$
also depends on the background counts. See LWP~7
in Table~\ref{t:detection} and \S\ref{s:lwp7}.}.
Since we only detected 9 periodic sources with \fap (Fix)$\le$ 1\%,
this result indicates there should be about 87$^{+17}_{-12}$ periodic sources,
which is 43$^{+9}_{-7}$\% of the 203 sources.  However, since the
detection probability drops significantly at net counts below 200,
we cannot make a reliable assessment of the faint periodic source
population. If we limit our detection probability at 1\%,
the dotted line in Fig.~\ref{detprob}b shows the effective population
of the sources we can explore. Accordingly the actual estimates of the
periodic population can be as low as the dotted green line in
Fig.~\ref{detprob}c. Therefore, in the worst case scenario
where most of the sources with net counts below 200 are not periodic,
the estimated periodic source population is 50$^{+10}_{-7}$, which 
is about 24$^{+5}_{-4}$\% of the hard X-ray sources.

Fig.~\ref{detprob}d shows the accumulated percentage of the
periodic sources in the LW as a function of net count.
Fig.~\ref{detprob}d shows a gradual increase of the fractional periodic
sources as net counts decrease to 200 from 500.  This observed variation
of the fractional periodic sources with net counts is consistent with
the fact that the X-ray fluxes of the majority
of MCVs are below $10^{32-33}$ erg s\sS{-1} \citep{Heinke08} since the
observed periodic sources are at the high end of the X-ray luminosity
distribution if they are in the Bulge.  It also implies the true
percentage of the periodic sources with net counts less than 200 is
likely higher than the value (43\%) at net count $\sim$ 200.

Table~\ref{simdet} summarizes the simulation results that estimate
the total number of periodic sources including unidentified ones among
the sources found in the LW.
We repeated the analysis for four sub-sets of the sources, which cover
samples only with non-CCD-edge sources
and hard X-ray sources ($E\Ss{50} \ge 2.5$ keV).  In this table, we
show the results for sources with net counts greater than 200, where
we have a sufficiently high detection probability of periodicity given
the number of sources.  The case using non-CCD-edge sources enables an
estimate of the systematic error in our selection procedure of periodic
sources since the observed periodicity of the sources that fall near a
CCD edge can be falsely discredited.  The case using only hard X-ray
sources should be a better representative of the Bulge sources since
they exclude the majority of the foreground soft sources.

A simple ratio argument between the observed periodic sources and
simulation results (column 6 \& 7 in Table~\ref{simdet}) indicates that
about 32--38\% of the sources in the LW should be, in fact, periodic
and the percentage increases to 43--51\% for the hard X-ray sources. The
errors of these estimates are calculated from the variances of the total
sample sizes (column 3) and the ranges of the detectable periodic sources
(column 5), provided  that at least the observed number
of periodic sources (column 4) are present in each case.  
Alternatively one can estimate the
errors for the total number of the periodic sources according to negative
binomial distribution
(column 8 \& 9 in Table~\ref{simdet}), which provides the variance estimates
of the unidentified (and thus total) number of periodic sources. 

The result indicates a large fraction of the Bulge X-ray sources
($\gtrsim$ 30--40\% of hard X-ray sources even with errors) found in
the GCR should be periodic. This provides a first direct evidence for
the presence of a large population of MCVs in the Bulge X-ray sources,
which supports that MCVs constitute a large majority of the 
Bulge X-ray sources.


\section{Summary} \label{s:sum}

We have found 10 periodic X-ray sources in the LW from
the 1 MS exposure of the \chandra ACIS-I instrument. 
The overall period distribution, X-ray luminosity and spectral properties
of these periodic sources and eight periodic sources found in the Sgr A*
field (M03b) fit the general description of MCVs, supporting the
argument that MCVs are the major constituent of the Bulge X-ray sources.
However, when inspecting the details of these X-ray properties -- period
distribution that resembles the polars' and the hard X-ray spectra that
resemble the IPs', we discover that these periodic sources may fit
into rare MCVs -- polars with unusally hard X-ray spectra or 
nearly synchronous MCVs.
The completeness tests indicate at least about 40\%
of the hard X-ray sources in the LW are periodic, which is a first direct
evidence of a large MCV population in the Bulge X-ray sources.  

Given the difficulty of identifying the nature of the Bulge X-ray
sources in the GCR due to high stellar density, large distance
and heavy obscuration, we recommend continuous X-ray monitoring of
the GCR for future discoveries of periodicity and unusual variability.
Considering the importance of these unusual MCVs in understanding the
evolutionary path of MCVs and their potential connection to thousands
of the Bulge X-ray sources, we also encourage systematic in-depth X-ray
studies of the similar kind (e.g.~there are only 13 ns-MCVs with \Ps/\Po $>3$
in the RK catalog) to establish a demographic profile of their X-ray
properties if any.

\section{Acknowledgment} 

We thank Dr.~Silas Laycock for the Magellan/IMACS observation and the
initial data processing of the LW.  This work is supported in part by
NASA/\chandra grants GO6-7088X, GO7-8090X and GO8-9093X. We also thank
Chandra X-ray Center (CXC) for their support in publication.



\begin{thebibliography}{}

\bibitem[Allen(1977)]{Allen77}
	Allen, K.~W., 1977,
	Astrophysical quantities., 3rd Ed., Moskva: Mir.

\bibitem[Anzolin(2008)]{Anzolin08}
	Anzolin, G., 2008,
	A\&A, 489, 1243.

\bibitem[Baganoff et al.(2003)]{Baganoff03}  
	Baganoff, F.~K.~et al. 2003,
	ApJ, 591, 891.

\bibitem[Bertin \& Arnouts(1996)]{Bertin96}  
	Bertin, E.~\& Arnouts, S.,~1996,
	ApJ, 117, 393.

\bibitem[Beuermann et al.(2009)]{Beuermann09}  
	Beuermann, K.~et al., 2009,
	A\&A, 507, 385.


\bibitem[Bretthorst(1988)]{Bretthorst88}
	Bretthorst, G.~L., 1988,
 	Bayesian Spectrum Analysis and Parameter Estimation,
	Springer-Verlag, Berlin Heidelberg
\bibitem[Buccheri et al.(1983)]{Buccheri83}  
	Buccheri, R.~et al. 1983,
	A\&A, 128, 245.

\bibitem[Butters et al.(2009)]{Butters09}  
	Butters, C.~H.~et al., 2009,
	A\&A, 498L, 17.

\bibitem[Casares et al.(1995))]{Casares95}  
	Casares, J.~et al., 1995, 
	MNRAS, 274, 565

\bibitem[Casares et al.(2011))]{Casares11}  
	Casares, J.~et al., 2011, 
	ATEL \#3206

\bibitem[Cleveland(1994)]{Cleveland94}  
	Cleveland, W.~S., 1994, 
	The Elements of Graphing Data, Hobart Press, 2nd edition.

\bibitem[Connors(2011)]{Connors11}
	Connors, A., 2011,
	in preparation.

\bibitem[Cummming(2002)]{Cumming02}  
	Cumming, A., 2002, 
	MNRAS, 333, 589

\bibitem[Degenaar \& Wijnands(2009)]{Degenaar09}  
	Degenaar, N.~\& R.~Wijnands, 2009,
	A\&A, 495, 547

\bibitem[Degenaar \& Wijnands(2010)]{Degenaar10}  
	Degenaar, N.~\& R.~Wijnands, 2010,
	A\&A, 524, 69

\bibitem[de Martino et al.(2006)]{Martino06}  
	de Martino, D.~et al., 2006,
	A\&A, 454, 287

\bibitem[Dolphin (2000)]{Dolphin00}  
	Dolphin, A.~E., 2000,
	PASP, 112, 1383.

\bibitem[Drimmel et al.(2003)]{Drimmel03}  
	Drimmel, R., Cabrera-Lavers, A., \&
	Lopez-Corredoira, M., 2003, 
	ApJ, 409, 205.

\bibitem[Evans \& Hellier (2004)]{Evans04a}  
	Evans, P.~A. \& Hellier, C., 2004,
	MNRAS, 353, 447.

\bibitem[Evans et al.(2004)]{Evans04b}  
	Evans, P.~A.~et al., 2004,
	MNRAS, 349, 715.

\bibitem[Evans \& Hellier (2005)]{Evans05}  
	Evans, P.~A. \& Hellier, C., 2005,
	ASP conference Series, 303, 2005.

\bibitem[Evans \& Hellier (2007)]{Evans07}  
	Evans, P.~A. \& Hellier, C., 2007,
	ApJ, 663, 1277.

\bibitem[Freeman et al.(2002)]{Freeman02}  
	Freeman, P.E.~et al. 2002,
	ApJS, 138, 185.

\bibitem[Gehrels et al.(1986)]{Gehrels86}  
	Gehrels, N.~et al. 1986,
	ApJ, 303, 336.

\bibitem[Grindlay et al.(2005)]{Grindlay05}  
	Grindlay, J.E.~et al., 2005,
	ApJ, 635, 907.

\bibitem[Grindlay et al.(2011)]{Grindlay11}  
	Grindlay, J.E.~et al., 2011,
	in preparation.

\bibitem[Heinke et al.(2008)]{Heinke08}
	Heinke, C.~O., et al., 2008,
	AIPC, 1010, 136.

\bibitem[Hellier, Beardmore \& Mukai(2002a)]{Hellier02a}  
	Hellier, C., Beardmore, A.~P., Mukai, K., 2002, 
	A\&A, 389, 904.

\bibitem[Hellier, Wynn \& Buckley(2002b)]{Hellier02b}  
	Hellier, C., Wynn, G.~A., Buckley, D.~A.~H., 2002, 
	MNRAS, 333, 84.

\bibitem[Hong et al.(2004)]{Hong04}  
 	Hong, J, Schlegel, E.~M.~\& Grindlay, J.~E., 2004,
	ApJ, 614, 508. (H04)

\bibitem[Hong et al.(2005)]{Hong05}  
 	Hong, J.~et al. 2005, 
	ApJ, 635, 907 (H05).

\bibitem[Hong et al.(2009a)]{Hong09a}  
 	Hong, J.~et al. 2009a, 
	ApJ, 699, 1053 (H09a).
\bibitem[Hong et al.(2009b)]{Hong09b}  
 	Hong, J.~et al. 2009b, 
	ApJ, 706, 223 (H09b).

\bibitem[Hong et al.(2011)]{Hong11}  
 	Hong, J, et al. 2011, 
	in preparation.

\bibitem[James et al.(2002)]{James02}  
	James, C.~H.~et al., 2002,
	MNRAS, 336, 550.

\bibitem[Koenig et al.(2008)]{Koenig08}  
	Koenig, X.~et al. 2008,
	ApJ, 685, 463.

\bibitem[Krimm, Kennea \& Holland(2011))]{Krimm11}  
	Krimm, H., Kennea, J.~A.~\& Holland S.~T., 2011, 
	ATEL \#3142

\bibitem[Laycock et al.(2005)]{Laycock05}  
	Laycock, S.~et al. 2005,
	ApJL, 634, 53 (L05).

\bibitem[Leahy et al.(1983)]{Leahy83}  
	Leahy, D.~A.~et al. 1983,
	ApJ, 266, 160 

\bibitem[Marsh \& Steeghs(2002)]{Marsh02}  
	Marsh, T.~R.~\& Steeghs, D., 2002
	MNRAS, 331L, 7

\bibitem[Marshall et al.(2006)]{Marshall06}
	Marshall, D. J.~et al. 2006, 
	A\&A, 453, 635 

\bibitem[Muno et al.(2003a)]{Muno03a}
	Muno, M. P.~et al. 2003a, 
	ApJ, 589, 225 (M03a)

\bibitem[Muno et al.(2003b)]{Muno03b}
	Muno, M. P.~et al. 2003b, 
	ApJ, 599, 465 (M03b).

\bibitem[Muno et al.(2004)]{Muno04}
	Muno, M. P.~et al. 2004, 
	ApJ, 613, 1179 (M04)

\bibitem[Muno et al.(2005)]{Muno05}
	Muno, M. P.~et al. 2005, 
	ApJL, 622, 113 

\bibitem[Muno et al.(2009)]{Muno09}
	Muno, M. P.~et al. 2009, 
	ApJS, 181, 110 (M09)

\bibitem[Norton et al.(1999)]{Norton99}  
	Norton, A.~J.~et al., 1999,
	A\&A, 347, 203

\bibitem[Norton et al.(2002)]{Norton02}  
	Norton, A.~J.~et al., 2002,
	A\&A, 384, 195

\bibitem[Norton, Wynn \& Somerscales(2004)]{Norton04}
	Norton, A.~J., Wynn, G.~A.~\& Somerscales, R.~V., 2004
	ApJ, 614, 349 

\bibitem[Norton et al.(2008)]{Norton08}
	Norton, A.~J., et al., 2008
	ApJ, 672, 524 

\bibitem[Parker, Norton, \& Mukai(2005)]{Parker05}  
	Parker, T.~J., Norton, A.~J., \& Mukai, K.,~2005,
  	A\&A 439, 213

\bibitem[Patterson(1998)]{Patterson98}  
	Patterson, J.,~1998,
  	PASJ, 110, 1132.

\bibitem[Pineault et al.(1987)]{Pineault87}
	Pineault, S., Landecker, T. L., \& Routledge, D., 1987, 
	ApJ, 315, 580

\bibitem[Predehl \& Schmitt(1995)]{Predehl95}
 	Predehl, P. \& Schmitt, J. H. M. M.,
	1995, A\&A, 293, 889.

\bibitem[Protassov(2002)]{Protassov02}  
	Protassov, R.,~et al.,~2002,
  	ApJ, 571, 545.

\bibitem[Ramsay et al.(2001)]{Ramsay01}
	Ramsay, G.~et al. 2001,
	MNRAS, 326L, 27.

\bibitem[Ramsay \& Cropper(2002)]{Ramsay02}  
	Ramsay, M. \& Cropper, M. ~2002,
  	MNRAS, 335, 918.

\bibitem[Ramsay \& Cropper(2002)]{Ramsay03}  
	Ramsay, M. \& Cropper, M. ~2003,
  	MNRAS, 338, 219.

\bibitem[Ramsay et al.(2004a)]{Ramsay04a}
	Ramsay, G.~et al. 2004a,
	MNRAS, 347, 95.

\bibitem[Ramsay et al.(2004b)]{Ramsay04b}
	Ramsay, G.~et al. 2004b,
	MNRAS, 350, 1373.

\bibitem[Ramsay et al.(2004c)]{Ramsay04c}
	Ramsay, G.~et al. 2004c,
	MNRAS, 354, 773.

\bibitem[Ramsay \& Cropper(2004)]{Ramsay04d}  
	Ramsay, M. \& Cropper, M. ~2004,
  	MNRAS, 347, 497.

\bibitem[Ramsay \& Cropper(2007)]{Ramsay07}  
	Ramsay, M. \& Cropper, M. ~2007,
  	MNRAS, 379, 1209.

\bibitem[Ramsay et al.(2008)]{Ramsay08}
	Ramsay, G.~et al. 2008,
	MNRAS, 387, 1157.

\bibitem[Rana et al(2004)]{Rana04}  
	Rana, V.~R.~et al., 2004,
	AJ, 127, 489.

\bibitem[Rana et al(2005)]{Rana05}  
	Rana, V.~R.~et al., 2005,
	ApJ, 625, 351.

\bibitem[Revnivtsev et al(2009)]{Revnivtsev09}  
	Revnivtsev, M., et al., 2009
	Nature, 458, 1142.

\bibitem[Revnivtsev et al(2010)]{Revnivtsev10}  
	Revnivtsev, M., et al., 2010
	A\&A, 515, 49.

\bibitem[Revnivtsev et al(2011)]{Revnivtsev11}  
	Revnivtsev, M., et al., 2011,
	astro-ph/1101.5883


\bibitem[Ritter \& Kolb(2003)]{Ritter03}
	Ritter, H. \& Kolb, U., 2003,
	ApJ, 404, 301 

\bibitem[Ruiter et al.(2006)]{Ruiter06}  
	Ruiter, A., Belczynski, K. \& Harrison, T.~2006,
  	ApJL, 640, 167.

\bibitem[Salinas \& Schlegel (2004)]{Salinas04}  
	Salinas, A.~\& Schlegel, E.~M., 2004,
	AJ, 128, 1331

\bibitem[Scargle(1982)]{Scargle82}  
	Scargle, J.~D., 1982,
  	ApJ, 263, 835.

\bibitem[Scargle(1998)]{Scargle98}  
	Scargle, J.~D., 1982,
  	ApJ, 504, 405.

\bibitem[Scaringi(2010)]{Scaringi10}  
	Scaringi, S., et al., 2010,
  	MNRAS, 401, 2207.

\bibitem[Schlegel et al.(1998)]{Schlegel98}  
	Schlegel, D., Finkbeiner, D. \& Davis, M., 1998,
  	ApJ, 500, 525.

\bibitem[Schlegel (2005)]{Schlegel05}  
	Schlegel, E.~M., 2005,
  	A\&A, 433, 635

\bibitem[Schmidt (2005)]{Schmidt05}  
	Schmidt, G.~D., 2005,
  	ApJ, 620, 422

\bibitem[Schodel et al.(2007)]{Schodel07}  
	Schodel, R.~et al., 2007,
  	A\&A, 469, 125

\bibitem[Schwarz et al.(2002)]{Schwarz02}  
	Schwarz, R., et al.,~2002,
  	A\&A 392, 505.

\bibitem[Schwarz et al.(2007)]{Schwarz07}  
	Schwarz, A.~D., et al.,~2007,
  	A\&A 473, 511.

\bibitem[Schwarz et al.(2005)]{Schwarz05}  
	Schwarz, R., et al.,~2005,
  	A\&A 442, 271.

\bibitem[Schwarz et al.(2009)]{Schwarz09}  
	Schwarz, R., et al.,~2009,
  	A\&A 496, 833.

\bibitem[Szkody et al.(2002)]{Szkody02}  
	Szkody, P.~et al., 2002,
	AJ, 123, 413.

\bibitem[Sumi(2004)]{Sumi04}  
	Sumi, T., 2004,
  	MNRAS, 349, 193.



\bibitem[Udalski(2002)]{Udalski02}  
	Udalski, T., 2002,
  	ACTA Astronomica, 52, 217.

\bibitem[van den Berg et al.(2006)]{Berg06}  
	van den Berg, M.~et al.. 2006, 
	ApJL, 135, 647.

\bibitem[van den Berg et al.(2009)]{Berg09}  
	van den Berg, M. et al. 2009, 
	ApJ, 700, 1702. (B09)

\bibitem[Vrielmann et al.(2004)]{Vrielmann04}  
	Vrielmann, S., Ness, J.-U., Schmitt, J.~H.~M.~M., 2004,
	4A\&A, 419, 673V.

\bibitem[Wang et al.(2002)]{Wang02}  
	Wang, Q.~D.~, Gotthelf, E.~V. \& Lang, C.~C., 2002,
	Nature, 415, 148.

\bibitem[Warner(1995)]{Warner95}  
	Warner, B., 1995,
	Cataclysmic Variable Stars, Cambridge Univ.~Press, New York

\bibitem[Wheatley \& West(2003)]{Wheatley03}  
	Wheatley, P.~J. \& West, R.~G., 2003,
  	MNRAS, 345, 1009.

\bibitem[Zhao et al.(2005)]{Zhao05}  
	Zhao, P.~et al. 2005,
	ApJS, 161, 429.

\end{thebibliography}
\end{document}